\documentclass{article}
\usepackage{jheppub} 
\usepackage[utf8x]{inputenc}	
\usepackage[english]{babel}

\usepackage[ddmmyyyy,hhmmss]{datetime}

\allowdisplaybreaks
\usepackage{amsxtra}
\usepackage{mathtools}
\usepackage{float}
\usepackage{tikz}
\usepackage{amsmath}
\usepackage{graphicx}
\usepackage{makecell}
\usepackage{mathrsfs}

\usepackage{tikz}
\usepackage{amsmath}
\usepackage{array}
\usetikzlibrary{decorations.pathreplacing, calc}

\usepackage{enumitem}

\usepackage{todonotes}
\usepackage[normalem]{ulem}

\title{Landscape of Higgs vacua in 5d $\mathcal{N}=1$}

\preprint{}

\author{Amihay Hanany, }
\author{Elias Van den Driessche}

\affiliation[]{Abdus Salam Centre for Theoretical Physics,\\ Imperial College London,\\ Prince Consort Road, SW7 2AZ, UK}

\abstract{Strongly coupled 5d $\mathcal{N}=1$ gauge theories have a Higgs branch of their moduli space of vacua, identified by 1/2-BPS operators and their algebraic relations. For $SU(N_c)_k$ theories, we map the moduli space across different phases and we establish how the Chern-Simons level $k$ deforms 
the chiral ring relations among mesons, baryons, and instantons, while determining specific patterns 
of symmetry enhancement. For $SO(N_c)$ theories, we construct the corresponding chiral ring 
constraints at infinite coupling. Finally, we identify the Higgs branch as a classical integrable system, through the expansion of the chiral ring in terms of action variables.}

\begin{document}
\maketitle
\flushbottom

\section{Introduction}
Since 5d $\mathcal{N}=1$ gauge theories are strongly coupled in the UV \cite{seiberg1996five}, they do not admit the usual Lagrangian description in terms of quarks and gluons. Nevertheless, the Higgs branch of their moduli space of vacua can still be defined through the \emph{chiral ring}, namely the ring of holomorphic\footnote{Holomorphic with respect to a given point on the $\mathbb{CP}^1$ worth of complex structures.} functions on the variety itself. Such ring is assumed to be in one to one correspondence with the 1/2-BPS operators of the theory, hence the name chiral, and is identified by two elements: a set of independent generators and a set of relations among them. The zero locus of such algebraic relations corresponds to the Higgs branch itself. \\

Through the program of \emph{magnetic quivers} (whose conceptual backbone is \cite{cabrera2019tropical}), elements of the chiral ring were enumerated for a multitude of theories, with different gauge algebras and matter couplings. Nonetheless, the effort to identify the independent elements and their relations is relatively recent and still ongoing \cite{cremonesi2017instanton, Hanany:2025jwo, Hanany:2026eyi}. In particular, this paper is going to explore the $SU(N_c)$ and $SO(N_c)$ gauge theories with matter in the fundamental representation, borrowing on the same techniques as \cite{Hanany:2025jwo}, which dealt instead with $Sp(N_c)$ theories. In particular, all chiral ring relations have been found using representation-theoretic techniques, namely \emph{Hilbert series} and the \emph{Plethystic} program \cite{Feng:2007ur}.\\

The identification of the 1/2-BPS fields and the relations among them, while of physical interest itself, allows to probe 5d Higgs branches with different instruments too. In particular it opens up the possibility, partially explored in \cite{Hanany:2026eyi}, to study the integrability of the chiral ring through Lagrangian submanifolds, as well as the deformation quantization thereof. The results in this paper are used in \cite{EVnew} to identify the integrable system, with its Lax matrix and spectral curve.\\

This paper heavily follows the notation and makes use of the results of \cite{cabrera2019tropical, Akhond:2024nyr}. The paper is organized as follows: in Sec.~\ref{Sec.SU} we will study the chiral ring of $SU(N_c)$ strongly coupled theories and its various phases, in Sec.~\ref{sec:SO} we will instead study the relatively simpler case of $SO(N_c)$ theories. Finally, in Sec.~\ref{sec:Lagrangian}, we will expand on the Lagrangian submanifolds of the Higgs branches of these theories.

\section{Chiral ring relations of SU theories}\label{Sec.SU}
\begin{figure}[h]
    \centering    \includegraphics[width=0.9\textwidth]{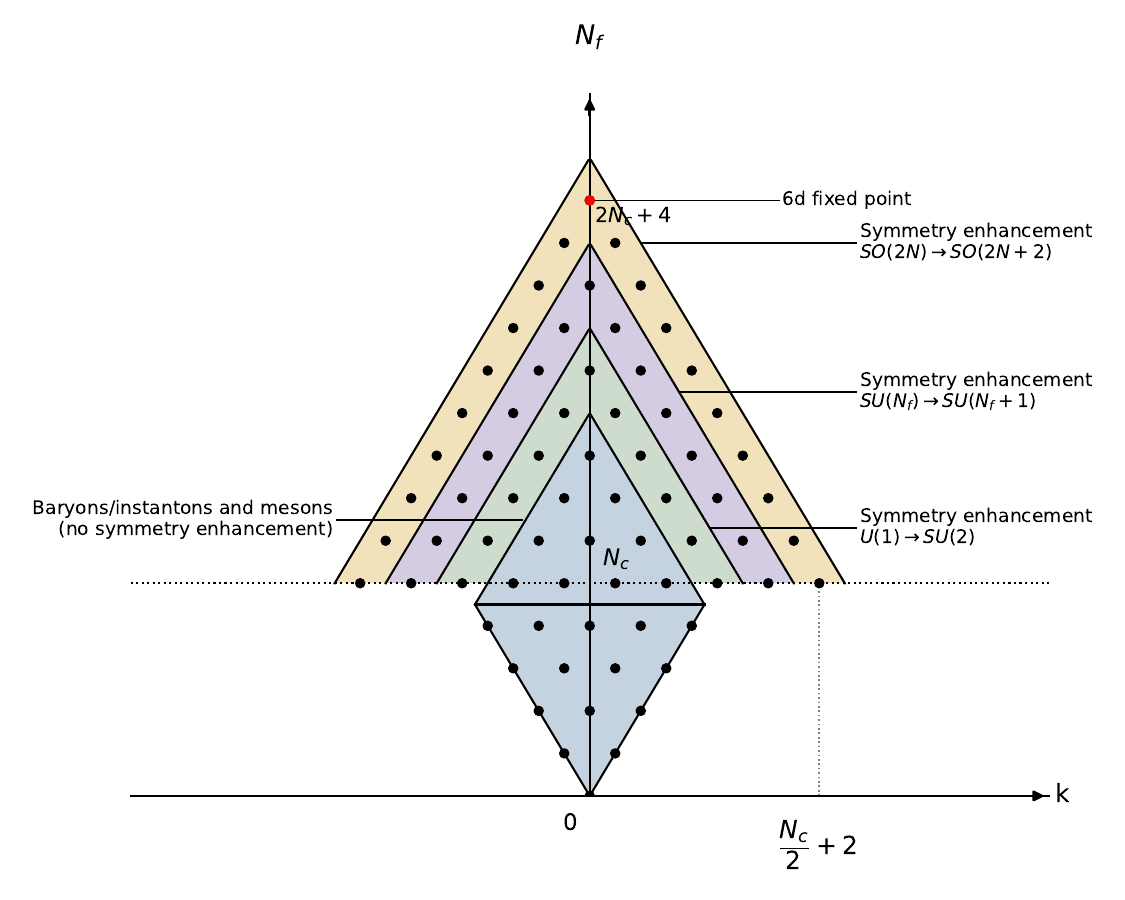}
    \caption{Phase diagram of strongly coupled $SU(N_c)$ theories, every black dot represents a different Higgs branch - the number of colours in the diagram is $N_c=5$. The $k$ level takes semi-integer values, while $N_f$ is integer, in compliance with $k+N_f/2\in \mathbb{Z}$. The diagram has $k\to -k$ reflection symmetry.}
    \label{fig:completephasediagram}
\end{figure}

In the finite coupling limit, the Higgs branch of $SU(N_c)$ theories with matter in the fundamental representation is independent \cite{seiberg1996five} of the Chern-Simons level $k$. Furthermore, depending on the number of colours $N_c$ and number of flavours $N_f$, the theory exhibits (see for example \cite{Bourget:2019rtl, Gray:2008yu, Argyres:1996eh}) two phases with different low energy spectra:
\begin{itemize}
    \item A purely mesonic branch, in which the scalars in the low energy theory are the moment maps\footnote{Moment map in the sense that they transform as primaries in the superconformal multiplet containing the conserved current of the flavour symmetry.} of the global symmetry, for any values of $(N_f,N_c)$.
    \item A baryonic branch, for $N_f \geq N_c$; in addition to the mesons, the spectrum also contains baryons obtained from the gauge-invariant anti-symmetrization of the quarks.
\end{itemize}
The two branches are distinct only for $N_c \leq N_f \leq 2N_c-2$, while they merge in a unique branch for higher $N_f$.\\
In the strongly coupled regime, there is a much more elaborate phase diagram of the Higgs branch due to the dependence on the Chern-Simons level, as represented in Fig.~\ref{fig:completephasediagram}.
We identify four regions (following the parameterization of \cite{cabrera2019tropical}):
\begin{itemize}
    \item Three-branched region: $|k| < N_c-N_f/2$; the low energy spectrum is composed by mesons and either baryons or instantons.
    \item Region II: $|k|=N_c-N_f/2$; this region is subject to symmetry enhancement of either or both the instanton and the baryon number. 
    \item Region III: $|k|=N_c-N_f/2+1$; this region is subject to symmetry enhancement of the classical flavour symmetry $SU(N_f)\times U(1)$ to $SU(N_f+1)$.
    \item Region IV: $|k|=N_c-N_f/2+2$; this region is subject to symmetry enhancement of $SU(N_f)\times U(1)_B \times U(1)_I$ to $SO(2N_f)\times SU(2)$ or $SO(2N_f+2)$, in the same way as $Sp(N_c)$ theories \cite{Hanany:2025jwo, Gaiotto:2015una}.
\end{itemize}
Each region is generally going to be composed of different phases. Different phases of the theory are identified by different sets of massless scalars; while multiple phases may coexist at given points of parameter space, they may be orthogonal to each other or overlapping in specific ways \cite{cabrera2019tropical}. For example, Region I has a baryonic phase and an instantonic phase, which intersect in a purely mesonic phase, as we are going to see.

\subsection{Three-branched region: $|k| < N_c-N_f/2$}
 The spectrum in this region includes mesons, baryons and instantons. While the baryonic invariants also exist in the IR, the presence of massless instantons in the spectrum is inherently a strong coupling effect. The chiral ring generators and their quantum numbers\footnote{The R-charge is the highest weight under $SU(2)_R$ rotations, while $U(1)_B$ denotes the baryon number and $U(1)_I$ the instanton number. The notation $\mu_1^a\mu_2^b\dots \mu_{N_f-1}^d$ is equivalent to the Dynkin labels $[a,b,\dots,d]$.} under global symmetries are reported in Tab.~\ref{tab:phaseIregIgen}.
    \renewcommand{\arraystretch}{1.5}
    \begin{table}[h]
        \centering
        \begin{tabular}{|c||c|c|c|c|c|c|} \hline
          Name  & Symbol & R-charge & \textbf{$SU(N_f)$} & \textbf{$U(1)_I$} & \textbf{$U(1)_B$} & Phases \\ \hline \hline
           Meson  &  $M_i^j$ & 2 & $\mu_1\mu_{N_f-1}$ & - & - & I, II, III \\ \hline
            Gaugino Bilinear & $S$ & 2 & 1 & - & - & I, II  \\ \hline \hline
           \makecell{Baryon \\ Anti-baryon} & \makecell{$B_{i_1\dots i_{N_c}}$\\ $\tilde{B}_{i_1\dots i_{N_f-N_c}}$} & $N_c$ & \makecell{$\mu_{N_c}$\\$\mu_{N_f-N_c}$} & - & $\pm 1$ & II \\ \hline \hline
           \makecell{Instanton \\ Anti-instanton} & \makecell{$I_{i_1\dots i_{a}}$\\ $\tilde{I}_{i_1\dots i_{N_f-a}}$} & $N_c$ & \makecell{$\mu_{a}$\\$\mu_{N_f-a}$} & $\pm 1$ & - & I \\ \hline
        \end{tabular}
        \caption{Spectrum on the three-branched region. The label $a$ represents $\dfrac{N_f}{2}-|k|$. The labelling is well defined since the instantonic phase exists for $2|k|\leq N_f < 2N_c-2|k|$, the baryonic phase exists for $N_c\leq N_f < 2N_c-2|k|$, while the mesonic phase exists for any $N_f \geq 2$ and $k > 1/2$. Furthermore, $k+N_f/2\in \mathbb{Z}$.}
        \label{tab:phaseIregIgen}
    \end{table}
 Let us notice that both baryons and instantons share the same R-charge, which however has different origins. For baryons, it is a consequence of the anti-symmetrization of quarks, while for instantons it comes from the dual Coxeter number of the gauge group. Let us also notice that the quantum number of the instantons are compatible with results obtained, with different techniques, in \cite{tachikawa2015instanton, Yonekura:2015ksa, zafrir2015instanton}.
 Furthermore, while the baryonic and instantonic phase may coexist in the same region of parameter space, they are not contiguous, as their intersection is a mesonic phase (cfr. Tab.~3 and 4 of \cite{cabrera2019tropical}). This implies that no theory in this region has, in its low energy spectrum, both baryons and instantons.\\
 We list the different types of constraints for each phase, derived with the same techniques of \cite{Hanany:2025jwo}. They all follow the same naturalness principle of \cite{Hanany:2025jwo}, according to which every relation that has the right quantum numbers to be corrected is indeed corrected in the UV. We start with the simplest phase, the one containing only mesons:

\paragraph{The mesonic phase} exists for $N_f\geq 2$ and $|k|>1/2$. It only contains mesons and it is indeed the closure of the maximal nilpotent orbit of height 2 of $SL(N_f,\mathbb{C})$, hence the unique constraint is part of the Joseph ideal \cite{joseph1976minimal} in Tab.~\ref{tab:Mnilpotent}.

\begin{table}[h]
    \centering
    \begin{tabular}{|c|c|c|} \hline
    Relation & R-charge & $SU(N_f)$  \\ \hline \hline
        $M_i^jM_j^k=0$ & 4 & $\mu_1\mu_{N_f-1}+1$ \\ \hline
    \end{tabular}
    \caption{Maximal nilpotent orbit constraints for the mesonic phase.}
    \label{tab:Mnilpotent}
\end{table}

\paragraph{The instantonic phase} sees the nilpotency of $M$ corrected by the gaugino bilinear. Furthermore, the instanton can covariantly combine with mesons and gauginos, as well as with the anti-instanton, as reported in Tab.~\ref{tab:constraintsphaseIregion1}.
\renewcommand{\arraystretch}{1.7}
\begin{table}[h]
    \centering
    \begin{tabular}{|c|c|c|c|} \hline
    Relation & R-charge & $SU(N_f)$ & $U(1)_I$ \\ \hline \hline
        $M_i^jM_j^k=S^2 \delta_i^k$ & 4 & $\mu_1\mu_{N_f-1}+1$ & 0 \\ \hline
        $M_{[i_1}^jI_{i_2\dots i_{a}]j}=SI_{i_1\dots i_a}$ & $N_c+2$ & $\mu_a$ & $1$ \\ \hline
        $M_{[i_1}^jI_{i_2\dots i_{a+1}]}\epsilon_{jp_1\dots p_{N_f-1}}=0$ & $N_c+2$ & $\mu_{a+1}\mu_{N_f-1}$ & $1$ \\  \hline
        $M_{i_1}^jI_{i_2\dots i_{a+1}}\epsilon_{jp_1\dots  p_{N_f-1}}\epsilon^{i_2\dots i_{a+1}p_1 \dots p_{N_f-a}}=0$ & $N_c+2$ & $\mu_{1}\mu_{N_f-a-1}$ & $1$ \\  \hline
        $I_{i_1 \dots i_{j} [i_{j+1} \dots i_a}\tilde{I}_{i_{a+1}\dots i_{N_f}]}=S^{N_c-j} M_{[i_1}^{p_1}\cdots M_{i_{j}]}^{p_{j}}\epsilon_{p_1\dots  p_{N_f}}$ & $2N_c$ & \makecell{$\mu_{j} \mu_{N_f-j}$\\ for $0 \leq j \leq a$} &  0 \\ \hline
        $I_{i_1 \dots i_{a-2j}[i_{a-2j+1}\dots a}I_{j_1 \dots j_a]}=0$ & $2N_c$ & \makecell{$\mu_{a-2j} \mu_{a+2j}$\\ for $0< j \leq \lfloor a/2 \rfloor$} & 2 \\ \hline
    \end{tabular}
    \caption{Chiral ring constraints for the instantonic phase. The $U(1)_I$-charged constraints are shown only for the instanton, since they are completely analogous for the anti-instanton. There may also be low rank accidents, where a relation is corrected in multiple ways or is missing.}
    \label{tab:constraintsphaseIregion1}
\end{table}
The last four relations in Tab.~\ref{tab:constraintsphaseIregion1} require projecting upon the required representation by subtraction of smaller representations (obtained through epsilon contraction). The same situation will appear throughout the paper; in order to avoid notation cluttering, we will always implicitly assume projection onto the irreducible representation appearing in the Plethystic logarithm.\\

The structure of the constraints is the same as the one described in \cite{Hanany:2025jwo}. The only meaningful difference is that the instantons transform here as differential forms, instead of spinors, allowing for the third and fourth constraints. In the theories examined in \cite{Hanany:2025jwo,Hanany:2026eyi}, the constraint:
\[
I\otimes \tilde{I}=S^{N_c}\exp(M/S)=\sum_{i=0}^{N_f}S^{N_c-i}\wedge^iM
\]
was shown to exist past the threshold of $R=2N_c$, however it carried no additional information. This was shown using the techniques of pure spinors, since for $Sp(N_c)$ theories the instantons transform as pure spinors of $SO(2N_f)$. Such techniques are lacking in the present case, hence we cannot make the same deductions. We can rewrite the relations in a more condensed way in Tab.~\ref{tab:condensedconstraintsphaseI}.

\renewcommand{\arraystretch}{2}
\begin{table}[h]
    \centering
    \begin{tabular}{|c|c|c|c|} \hline
    Relation & R-charge & $SU(N_f)$ & $U(1)_I$ \\ \hline \hline
        $M^2=S^2$ & 4 & $\mu_1\mu_{N_f-1}+1$ & 0 \\ \hline
        $M \times I=SI$ & $N_c+2$ & $\mu_a+\mu_{a+1}\mu_{N_f-1}+\mu_{1}\mu_{N_f-a-1}$ & $1$ \\ \hline
        $I\otimes \tilde{I}=S^{N_c}\exp_a(M/S)$ & $2N_c$ & $\sum_{i=0}^a\mu_i \mu_{N_f-i}$ & 0 \\ \hline
        $\text{Sym}^2I =(II)_{\mu_{a}^2}$ & $2N_c$ & $\mu_{a-2}\mu_{a+2},\mu_{a-4}\mu_{a+4}\dots$ & 2 \\ \hline
    \end{tabular}
    \caption{Condensed form of the chiral ring constraints for the instantonic phase. Henceforth we will use the short-hand notation of the second line, with the symbol $\times$ denoting three different operations: the contraction with $\delta$ and upstairs and downstairs contraction with $\epsilon$. As previously, $a=N_f/2-|k|$.}
    \label{tab:condensedconstraintsphaseI}
\end{table}
for ease of notation we have defined the exponential of a 2-form $x$ up to power $a$:
\begin{equation}
    \exp_a(x)=\sum_{i=0}^{a}\wedge^{i}(x)=1+x+\wedge^2x+\dots + \wedge^{a}x \ .
\end{equation}

The last relation in Tab.~\ref{tab:condensedconstraintsphaseI} is reminiscent of the pure spinor condition examined in \cite{Hanany:2025jwo, Hanany:2026eyi}. As in that case, it appears for $N_f\geq 4$ and constrains the symmetric product of the instanton representation to be equal to the highest component only. It's easy to see if we make an example; for $N_f=5$, taking a instanton in the $\mu_2$, the symmetric product reads:
\[
\text{Sym}^2\mu_2=\mu_2^2+\mu_4
\]
and the purity condition implies that the $\mu_4$ representation vanishes, or in tensor notation:
\[
I_{[a_1a_2}I_{a_3a_4]}=0
\]
In all generality, the symmetric product of the instanton representation is:
\begin{equation}
 \text{Sym}^2\mu_{a}= 
  \mu_a^2+\mu_{a-2}\mu_{a+2}+\mu_{a-4}\mu_{a+4}\dots
\end{equation}
and the purity condition implies that all representations vanish except the first one.\\

\begin{figure}[h]
    \centering
    \includegraphics[width=0.70\textwidth]{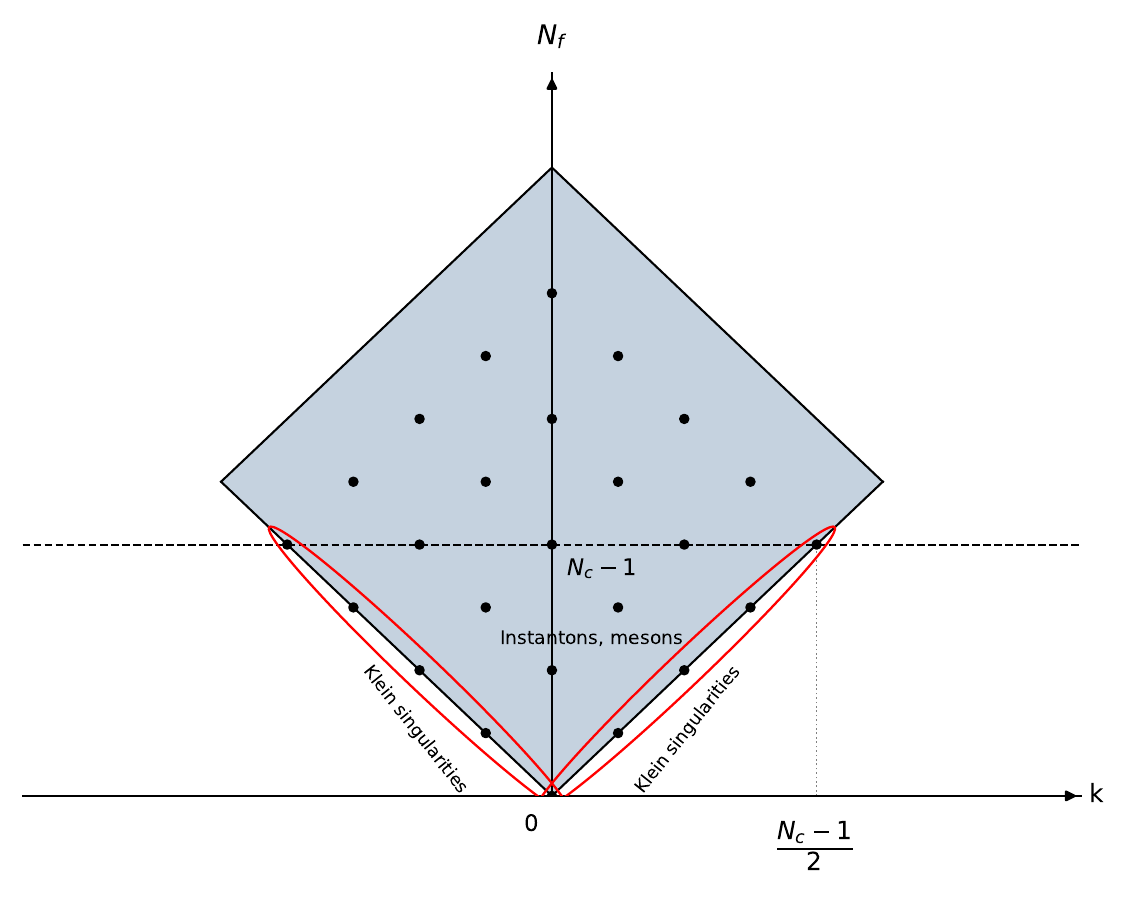}
    \caption{Instantonic phase of strongly coupled $SU(N_c)$ theories, every black dot represents a different Higgs branch - the number of colours in the diagram is $N_c=5$. The $k$ level takes semi-integer values, while $N_f$ is integer, in compliance with $k+N_f/2\in \mathbb{Z}$. The diagram has $k\to -k$ reflection symmetry. Only phases with generators of R-charge $>2$ are shown; mesonic Phase III is not denoted.}
    \label{fig:phase1}
\end{figure}

The constraints in Tab.~\ref{tab:condensedconstraintsphaseI} hold for the theories in the region of parameter space drawn in Fig.~\ref{fig:phase1}. We remark in particular that, while the pure theory is known \cite{cremonesi2017instanton} to have a $A_n$ Klein singularity (of type $\mathbb{C}^2/\mathbb{Z}_{N_c}$) as its Higgs branch, this is indeed a feature of a whole line of flavoured theories with nonvanishing $k$ level. 

As an example, let us consider the Higgs branch of $SU(3)_{1/2}$ with $3$ flavours. The Hilbert series\footnote{It is not particularly relevant, for our purposes, to know each and every term of the Hilbert series. That is why some terms are left dotted: we only care about the elements that appear, order by order, both in the Hilbert series and the Plethystic logarithm.} thereof is:
\renewcommand{\arraystretch}{1.2}
 \begin{equation}
            \begin{array}{ll}
               \text{HS}(t)  & =  1+(1+\mu_1\mu_2)t^2+\\
                 & +(q\mu_1+q^{-1}\mu_2)t^3+\\
                 &+(1+\mu_1\mu_2+\mu_1^2\mu_2^2)t^4+ \\
                 & + (q\mu_1+q^{-1}\mu_2+\dots)t^5+\\
                 & + (1+\mu_1\mu_2+\dots)t^6
            \end{array}
        \end{equation}
        and the Plethystic logarithm is:
        \renewcommand{\arraystretch}{1.2}
         \begin{equation}
            \begin{array}{ll}
                \text{PL}(t) & = (1+\mu_1\mu_2)t^2+ \\
                 & +(q\mu_1+q^{-1}\mu_2)t^3+\\
                 & -(1+\mu_1\mu_2)t^4+\\& - (q(\mu_2^2+\mu_1)+q^{-1}(\mu_1^2+\mu_2))t^5 +\\
                 & -t^6+\dots \\
            \end{array}
        \end{equation}
         hence the generators are reported in Tab.~\ref{tab:SU(3)1/2+3,gen} and the relations in Tab.~\ref{tab:tab:SU(3)1/2+3,rel}.
        \renewcommand{\arraystretch}{1.3}
        \begin{table}[h]
            \centering
            \begin{tabular}{|c|c|c|c|} \hline 
                & R-charge & $SU(N_f)$ & $U(1)_I$ \\ \hline \hline
               $M_i^j$  & 2 &  $\mu_1\mu_2$ & 0 \\ \hline
               $S$ & 2 & 1 & 0 \\ \hline
               $I_{i}$ & 3 & $\mu_1$ & $1$\\ \hline
               $\tilde{I}_{ij}$ & 3 & $\mu_2$ & $-1$\\ \hline
            \end{tabular}
            \caption{Chiral ring generators of strongly coupled $SU(3)_{1/2}+3F$.}
            \label{tab:SU(3)1/2+3,gen}
        \end{table}
      
        \begin{table}[h]
            \centering
            \begin{tabular}{|c|c|c|c|} \hline
                 & R-charge & $SU(N_f)$ & $U(1)_I$  \\ \hline \hline
                $M_i^jM_j^k=S^2\delta_i^k$ & 4 & $\mu_1\mu_2+1$ & 0 \\ \hline
                $I_{j}M_{r}^{j}=I_{r}S$ & 5 & $\mu_1$ & $1$  \\ \hline
                $I_{[i}M_{r]}^{j}\epsilon_{ja_1 a_2}=0$ & 5 & $\mu_2^2$ & $ 1$  \\ \hline
                 $I_i \tilde{I}_{jk}\epsilon^{ijk}=S^3$ & 6 & $1$ & 0 \\ \hline
                 $I_i\tilde{I}_{jk}=S^2M_{i}^p\epsilon_{pjk}$ & 6 & $\mu_1\mu_2$ & 0 \\ \hline
            \end{tabular}
            \caption{Chiral ring relations of strongly coupled $SU(3)_{1/2}+3$. This is a relatively simple example since it has low $N_c$ and some of the relations in Tab.~\ref{tab:constraintsphaseIregion1} are suppressed due to low rank of the flavour group; an example at higher rank is worked out in \S\ref{par:firstexample}. The constraint in line 3 is implicitly assumed to be symmetrized in the two pairs of indices.}
            \label{tab:tab:SU(3)1/2+3,rel}
        \end{table}

\paragraph{The baryonic phase} is populated by mesons and baryons; while classically baryons are independent of the Chern-Simons level, in the UV limit baryons are present without symmetry enhancement only for $N_c \leq N_f < 2N_c-2|k|$. Essentially, the chiral ring relations are the same as Tab.~\ref{fig:phase1}, since the only difference between baryons and instantons is the representation under flavour symmetry.
Again we can condense the form of the chiral ring constraints in Tab.~\ref{tab:constraintsphaseII}. Strikingly, the relations involving the baryons are not the classical relations often presented. Indeed, the presence of the gaugino bilinear allows for corrections. Since these are supposed to be classical objects, it should be possible in principle to reproduce the chiral ring relations from a Lagrangian. It remains an open problem to do so.
\renewcommand{\arraystretch}{1.7}
\begin{table}[h]
    \centering
    \begin{tabular}{|c|c|c|c|} \hline
    Relation & R-charge & $SU(N_f)$ & $U(1)_I$ \\ \hline \hline
        $M^2=S^2$ & 4 & $\mu_1\mu_{N_f-1}+1$ & 0 \\ \hline
        $M \times B=SB$ & $N_c+2$ & $\mu_{N_c}+\mu_1\mu_{N_f-N_c-1}+\mu_{N_c+1}\mu_{N_f-1}$ & $1$ \\ \hline
        $B\otimes \tilde{B}=S^{N_c}\exp_{N_c}(M/S)$ & $2N_c$ & $\sum_{j=0}^{N_c}\mu_j \mu_{N_f-j}$ & 0 \\ \hline
        $\text{Sym}^2B=(BB)_{\mu_{N_c}^2}$ & $2N_c$ & $\mu_{N_c-2}\mu_{N_c+2},\dots$ & 2 \\ \hline
    \end{tabular}
    \caption{Condensed form of the chiral ring constraints for the baryonic phase. An example is worked out in \S \ref{par:baryonexample}.}
    \label{tab:constraintsphaseII}
\end{table}

\begin{figure}[h]
    \centering    \includegraphics[width=0.70\textwidth]{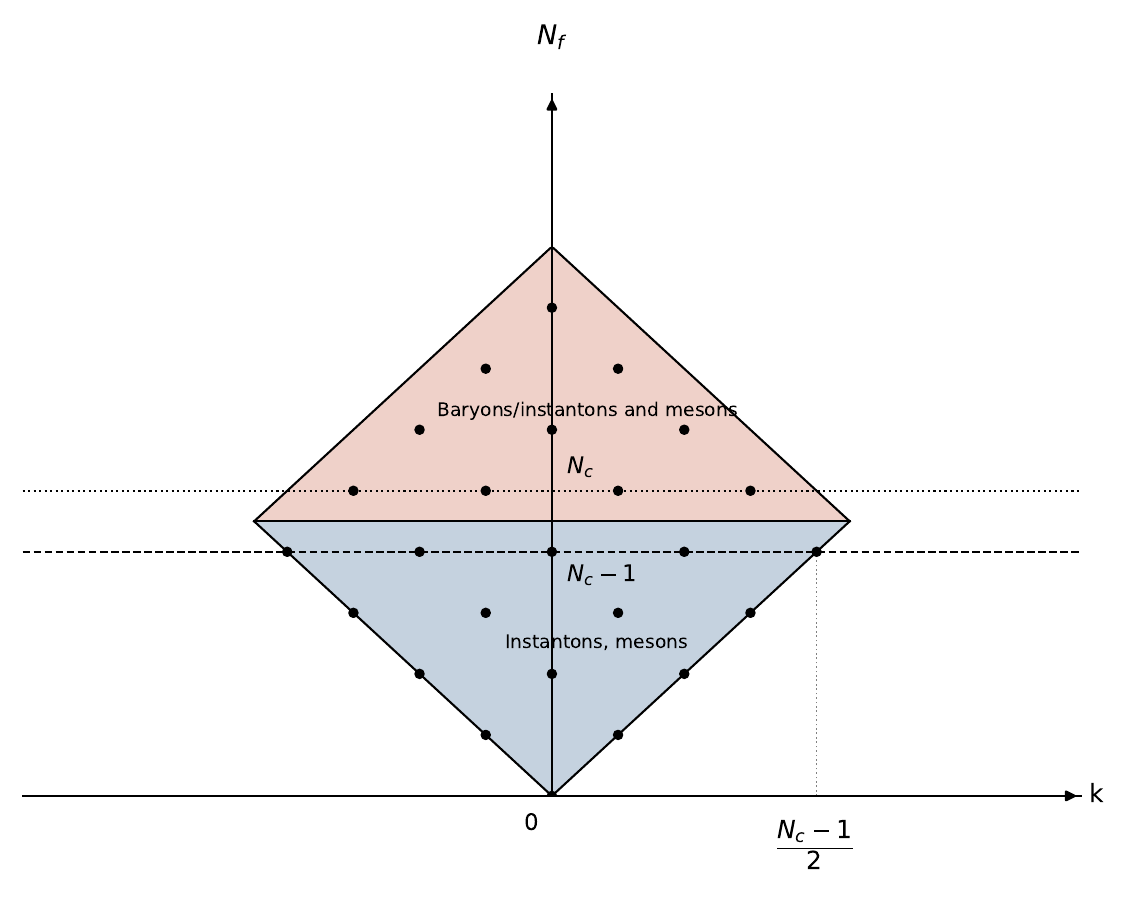}
    \vspace{-0.2cm}
    \caption{Baryonic and instantonic phases of strongly coupled $SU(N_c)$ theories, every black dot represents a different Higgs branch - the number of colours in the diagram is $N_c=5$. The $k$ level takes semi-integer values, while $N_f$ is integer, in compliance with $k+N_f/2\in \mathbb{Z}$. Only phases with generators of R-charge $R>2$ are shown.}
    \label{fig:phase2}
\end{figure}

The complete phase diagram is reported in Fig.~\ref{fig:phase2}.
\vspace{-0.3cm}

\subsection{Region II: $|k|=N_c-N_f/2$}
\begin{figure}[h]
    \centering
\includegraphics[width=0.7\textwidth]{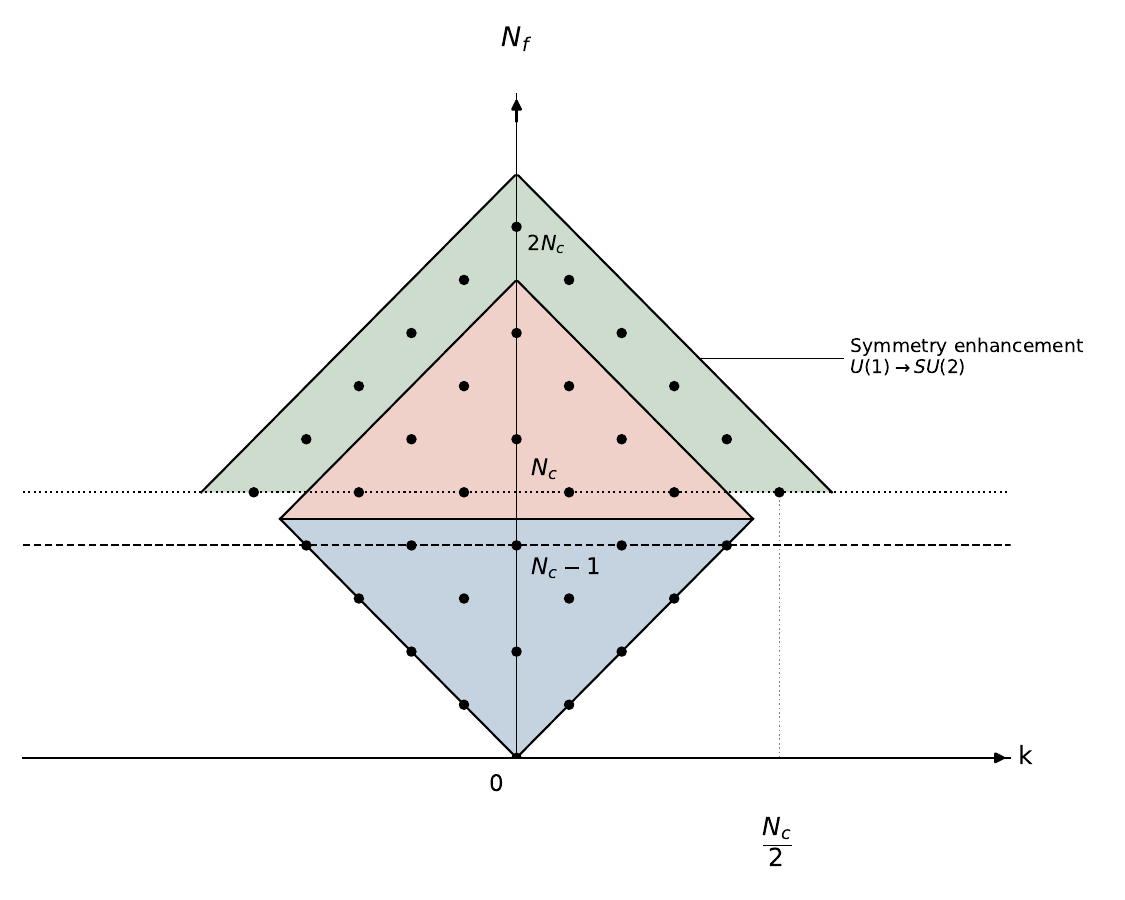}
    \caption{Phase diagram of strongly coupled $SU(N_c)_{k}+N_f$, including the region of partial symmetry enhancement. Only the phases with generators of R-charge $R>2$ are shown. The reference value of $N_c$ in the diagram is $N_c=5$.}
    \label{fig:region2}
\end{figure}

This region is identified by $|k|=N_c-N_f/2$ and sees partial enhancement of the global symmetry. In particular the $U(1)_I$ and $U(1)_B$ are either separately or both enhanced to $SU(2)$. The phase diagram is reported in Fig.~\ref{fig:region2}.

The pattern of symmetry enhancement (cfr. \S 5.2 of \cite{cabrera2019tropical}) depends on the value of $k$:
\begin{itemize}
    \item For $k=0$, symmetry is enhanced from $SU(N_f)\times U(1)_I \times U(1)_B$ to $SU(N_f)\times SU(2) \times SU(2)$. There is a unique phase of the theory, and the generators of the chiral ring in this case are reported in Tab.~\ref{tab:genSU(Nc)+2Nc} and the relations in Tab.~\ref{tab:region2}.
 \renewcommand{\arraystretch}{1.3}
        \begin{table}[h]
            \centering
            \begin{tabular}{|c|c|c|c|c|} \hline 
                & R-charge & $SU(N_f)$ & $SU(2)_{\nu}$ & $SU(2)_{\rho}$  \\ \hline \hline
               $M_i^j$  & 2 &  $\mu_1\mu_{N_f-1}$ & 1 & 1 \\ \hline
               $S^{(ab)}$ & 2 & 1 & $\nu^2$ & 1 \\ \hline
                $S^{(\dot{a}\dot{b})}$ & 2 & 1 & 1 & $\rho^2$ \\ \hline
               $B_{i_1\dots i_{N_f/2}}^{a,\dot{a}}$ & $N_c$ & $\mu_{N_c}$ & $\nu$ & $\rho$\\ \hline
            \end{tabular}
            \caption{Generators of the chiral ring of strongly coupled $SU(N_c)_0+2N_c$.}
            \label{tab:genSU(Nc)+2Nc}
        \end{table}
    The spectrum differs from the one in the IR for the mixing of $U(1)_I$ and $U(1)_B$ into two $SU(2)$ groups. The representation of the instantons is the same as the baryons' under $SU(N_f)$, suggesting that the generator $B$ is a combination thereof. In particular, the weights of $B$ under the two $SU(2)$ groups correspond to combinations of the classical baryons, anti-baryon, instanton and anti-instanton. This is made possible by their shared representation under $SU(N_f)$.
    \renewcommand{\arraystretch}{2.2}
    \begin{table}[h]
            \centering
            \begin{tabular}{|c|c|c|c|c|} \hline 
                & R-charge & $SU(N_f)$ & $SU(2)_{\nu}$ & $SU(2)_{\rho}$  \\ \hline \hline
               $M^2=(S^{ab})^2=(S^{\dot{a}\dot{b}})^2$ & 4 & $\mu_1\mu_{N_f-1}+2$ & 1 & 1 \\ \hline
               \makecell{$M \times B^{a,\dot{a}}=B^{b,\dot{a}}S^{ab}=B^{a,\dot{b}}S^{\dot{a}\dot{b}}$} & $N_c+2$ & \makecell{$2\mu_{N_c}+\mu_1\mu_{N_c-1}+$\\$+\mu_{1+N_c}\mu_{N_f-1}$} & $\nu$ & $\rho$ \\ \hline
                
$B\otimes B= S\dot{S}\sum_{i=1}^{\lfloor N_c/2 \rfloor }\text{tr}M^{2i-2}*\wedge^{N_c-2i}M$ & $2N_c$ & $\mu_{N_c-2i}\mu_{N_c+2i}$ & $\nu^2$ & $\rho^2$ \\ \hline

$B\otimes B=S \sum_{i=0}^{\lfloor (N_c-1)/2 \rfloor}\text{tr}M^{2i}\wedge^{N_c-1-2i}M$ & $2N_c$ & \makecell{$\mu_{N_c-1-2i}\mu_{N_c+1+2i}$} & $\nu^2$ & 1 \\ \hline

$B\otimes B=\dot{S} \sum_{i=0}^{\lfloor (N_c-1)/2 \rfloor}\text{tr}M^{2i}\wedge^{N_c-1-2i}M$ & $2N_c$ & \makecell{$\mu_{N_c-1-2i}\mu_{N_c+1+2i}$} & $1$ & $\rho^2$ \\ \hline
$B\otimes B= *\wedge^{N_c}M$ & $2N_c$ & $\mu_{N_f/2}^2$ & $1$ & $1$ \\ \hline
 
            \end{tabular}
            \caption{Chiral ring relations for region II of strongly coupled $SU(N_c)_0+(2N_c)F$. An example is worked out in \S \ref{par:secondexample}. It is worth remembering that odd Casimirs of $M$ vanish. Contraction with the $\epsilon$ invariant of $SU(2)$ is assumed.}
            \label{tab:region2}
        \end{table}

\item For $k = 1/2$, symmetry is enhanced to $SU(N_f)\times SU(2)\times U(1)$. Again there is a unique phase of the theory, and the generators of the chiral ring are reported in Tab.~\ref{tab:region2-2}, and the relations in Tab.~\ref{tab:region2bigrel}.
      \renewcommand{\arraystretch}{1.3}
        \begin{table}[h]
            \centering
            \begin{tabular}{|c|c|c|c|c|} \hline 
                & R-charge & $SU(N_f)$ & $SU(2)$ & $U(1)$  \\ \hline \hline
               $M_i^j$  & 2 &  $\mu_1\mu_{N_f-1}$ & 1 & 0 \\ \hline
               $S$ & 2 & 1 & 1 & 0 \\ \hline
                $T^{(\alpha\beta)}$ & 2 & 1 &   $\nu^2$ & 0 \\ \hline
               $B_{i_1\dots i_a}^{\alpha}$ & $N_c$ & $\mu_{a}$ & $\nu$ &1 \\ \hline
                $\tilde{B}_{i_1\dots i_{N_f-a}}^{\alpha}$ & $N_c$ & $\mu_{N_f-a}$ & $\nu$ & $-1$\\ \hline
            \end{tabular}
            \caption{Chiral ring generators for $SU(N_c)_{1/2}+(2N_C-1)F$ at strong coupling. The label $a$ represents, as previously, $a=-k+N_f/2$.}
            \label{tab:region2-2}
        \end{table}
     \renewcommand{\arraystretch}{2}
         \begin{table}[h]
            \centering
            \begin{tabular}{|c|c|c|c|c|} \hline 
                & R-charge & $SU(N_f)$ & $SU(2)$ & $U(1)$  \\ \hline \hline
            $M^2=S^2=T^2$ & 4 & $\mu_1\mu_{N_f-1}+2$ & 1 & 0 \\ \hline
            $
            B\times M=SB=TB$ & $N_c+2$ & \makecell{$2\mu_a+\mu_1\mu_{N_f-a-1}+$\\$+\mu_{a+1}\mu_{N_f-1}$} & $\nu+1$ & 1 \\ \hline

    \makecell{$B \otimes \tilde{B}=$\\$=T\sum_{i=0}^{a}S^{N_c-i-a-1}*\wedge^{a-i}M$} & $2N_c$ &  \makecell{$\mu_{a-i}\mu_{N_f-a+i}$} &  $\nu^2$ & 0 \\ \hline
               \makecell{$B\otimes \tilde{B}=$\\$=\sum_{i=0}^{\lfloor (a-1)/2 \rfloor}S^{N_c-a+2i} * \wedge^{a-2i} M$} & $2N_c$ & \makecell{$\mu_{a-2i}\mu_{N_f-a+2i}$} & $1$ & 0 \\ \hline
               $\text{Sym}^2B=(BB)_{\mu_a^2}$ & $2N_c$ & $\mu_{a-2}\mu_{a+2},\dots$ & $\nu^2$ & 2 \\ \hline
               $\wedge^2B=*(BB)$ & $2N_c$ & $\mu_{a-1}\mu_{a+1},\mu_{a-3}\mu_{a+3},\dots$ & 1 & 2 \\ \hline
            \end{tabular} 
            \caption{Chiral ring relations for $SU(N_c)_{1/2}+(2N_c-1)F$ at strong coupling. The label $a$ represents, as previously, $a=-|k|+N_f/2$. An example is worked out in \S \ref{sec:1/2example}.}
            \label{tab:region2bigrel}
        \end{table} 

    \item For $k=1$ there are two phases of the theory:
    \begin{itemize}
        \item Phase I exists for $N_f/2 \geq |k|$ and has enhancement to $SU(N_f)\times SU(2) \times U(1)$. The chiral ring generators and relations are the same as the ones, respectively, in Tab.~\ref{tab:region2-2} and Tab.~\ref{tab:region2bigrel}.
        \item Phase II, for any $N_f \geq 1$, has no enhancement, and is the closure of the maximal nilpotent orbit (of height 2) of $SL(N_f,\mathbb{C})$.
    \end{itemize}
    \item For $k>1$, there are two phases of the theory:
    \begin{itemize}
        \item Phase I, for $N_f/2 \geq |k|$, has enhancement to $SU(N_f)\times SU(2) \times U(1)$. The chiral ring generators and relations are the same as the ones, respectively, in Tab.~\ref{tab:region2-2} and Tab.~\ref{tab:region2bigrel}.
        \item Phase II, for any $N_f \geq 1$, has enhancement to $SU(N_f)\times SU(2)$, the chiral ring generators are reported in Tab.~\ref{tab:gensphaseIIregionII} and the relations in Tab.~\ref{tab:relsphaseIIregionII}.
    \renewcommand{\arraystretch}{1.3}
        \begin{table}[h]
            \centering
            \begin{tabular}{|c|c|c|c|} \hline 
                & R-charge  & $SU(N_f)$ & $SU(2)$  \\ \hline \hline
               $M_i^j$  & 2 &  $\mu_1\mu_{N_f-1}$ & 0 \\ \hline
               $S^{(ab)}$ & 2  & $\nu^2$ & 0 \\ \hline
            \end{tabular}
            \caption{Chiral ring generators of the Phase II, in region II, of $SU(N_c)_k+N_f$.}
            \label{tab:gensphaseIIregionII}
        \end{table}
 \begin{table}[h]
            \centering
            \begin{tabular}{|c|c|c|c|} \hline 
                & R-charge  & $SU(N_f)$ & $SU(2)$  \\ \hline \hline
               $M^2=0$ & 4 & $\mu_1\mu_{N_f-1}+1$ & 1 \\ \hline
               $S^2=0$ & 4 & $1$ & 1 \\ \hline
            \end{tabular}
            \caption{Chiral ring relations of the Phase II, in region II, of $SU(N_c)_k+N_f$.}
            \label{tab:relsphaseIIregionII}
        \end{table}        
    \end{itemize}
\end{itemize}

\subsection{Region III: $|k|=N_c-N_f/2+1$}
\begin{figure}[h]
    \centering
    \includegraphics[width=0.8\textwidth]{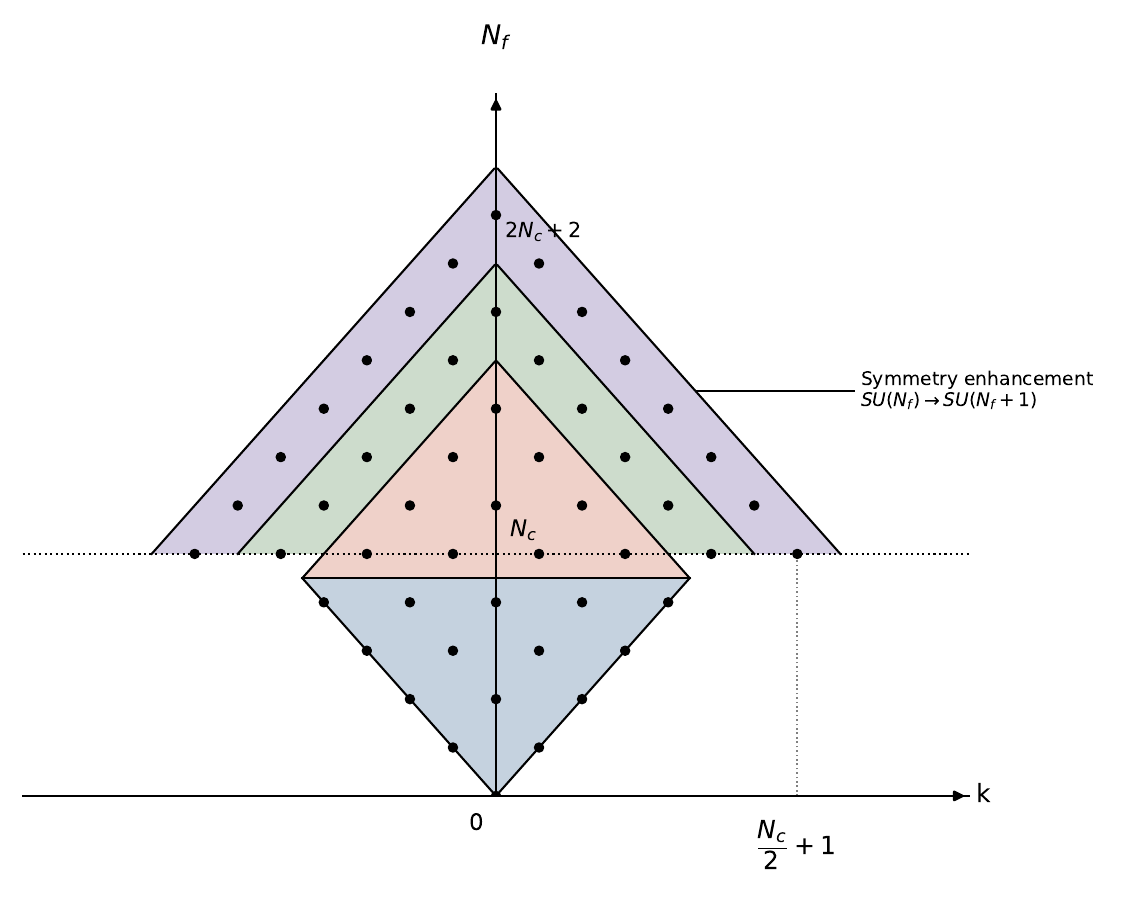}
    \caption{Phase diagram of strongly coupled $SU(N_c)_{k}+N_f$, including the third region with $SU(N_f)\times U(1)\to SU(N_f+1)$. Only the phases with generators of R-charge $R>2$ are shown. The reference value of $N_c$ in the diagram is $N_c=5$.}
    \label{fig:region3}
\end{figure}

The third region (cf. \S 5.3 of \cite{cabrera2019tropical}) is defined for $|k|=N_c+1-N_f/2$ and has symmetry enhancement of the type $SU(N_f)\times U(1)\to SU(N_f+1)$. The phase diagram is reported in Fig.~\ref{fig:region3}.

The pattern of symmetry enhancement again depends on the Chern-Simons level:
\begin{itemize}
    \item Phase I exists only for $k=0$, where the symmetry is enhanced from $SU(N_f)\times U(1)_I\times U(1)_B$ to $SU(N_f+2)$.  The chiral ring generators are reported in Tab.~\ref{tab:region3phase1gen} and the relations in Tab.~\ref{tab:region3phase1rel}.
       \renewcommand{\arraystretch}{1.3}
        \begin{table}[h]
            \centering
            \begin{tabular}{|c|c|c|} \hline 
                & R-charge  & $SU(N_f+2)$ \\ \hline \hline
               $M_i^j$  & 2 &  $\mu_1\mu_{N_f+1}$ \\ \hline
               $I_{i_1\dots i_{N_c+2}}$ & $N_c$ & $\mu_{N_c+2}$ \\ \hline
            \end{tabular}
            \caption{Chiral ring generators of $SU(N_c)_0+(2N_c+2)F$ at strong coupling (Region III).}
            \label{tab:region3phase1gen}
        \end{table}
    \renewcommand{\arraystretch}{1.3}
        \begin{table}[h]
            \centering
            \begin{tabular}{|c|c|c|} \hline 
              Constraints  & R-charge  & $SU(N_f+2)$ \\ \hline \hline
             $M^2=0$ & 4 & $\mu_1\mu_{N_f+1}$ \\ \hline
             $(M  I)\epsilon=0$  & $N_c+2$ & $\mu_1\mu_{N_f/2}+\mu_{N_f/2+2}\mu_{N_f+1}$ \\ \hline
              $I \otimes I=\wedge^{N_c}M$ & $2N_c$ & $\mu_{N_c} \mu_{N_f+2-N_c}$ \\ \hline
            \end{tabular}
            \caption{Chiral ring relations of strongly coupled $SU(N_c)_{0}+(2N_c+2)F$ (regione III).}
            \label{tab:region3phase1rel}
        \end{table}
An example is worked out in \S\ref{sec:0example}. The last constraint in Tab.~\ref{tab:region3phase1rel} does not exhibit the usual series of powers since there is no $S$ to absorb the R-charge. Furthermore, the meson matrix $M$ in the adjoint of $SU(N_f+2)$ combines both the classical meson matrix, the gaugino bilinear and instantons of R-charge equal to 2. This is analogous to what happens for $Sp(1)+7F$, as studied in \cite{cremonesi2017instanton}.
\item Phase II, exists only for $k=1/2$, where the symmetry is enhanced to $SU(N_f+1)\times SU(2)$. The chiral ring generators are Tab.~\ref{tab:region3phase2gen} and the relations in Tab.~\ref{tab:region3phase2rel}.
      \renewcommand{\arraystretch}{1.3}
        \begin{table}[h]
            \centering
            \begin{tabular}{|c|c|c|c|} \hline 
                & R-charge  & $SU(N_f+1)$ & $SU(2)$ \\ \hline \hline
           $M_i^j$  & 2 &  $\mu_1\mu_{N_f}$ & 1 \\ \hline
               $S^{(ab)}$ & 2 & 1 & $\nu^2$ \\ \hline
               $B_{i_1\dots i_{N_c+1}}^a$ & $N_c+1$ & $\mu_{N_c+1}$ & $\nu$ \\ \hline
            \end{tabular}
            \caption{Chiral ring generators of strongly coupled $SU(N_c)_{1/2}+(2N_c+1)F$.}
            \label{tab:region3phase2gen}
        \end{table}
         \renewcommand{\arraystretch}{1.5}
        \begin{table}[h]
            \centering
            \begin{tabular}{|c|c|c|c|} \hline 
                & R-charge  & $SU(N_f+1)$ & $SU(2)$ \\ \hline \hline
               $M^2=S^2$  & 4 &  $\mu_1\mu_{N_f}+1$ & 1 \\ \hline
      $B \times M=BS$ &  $N_c+3$ & $\mu_{N_c+1},\mu_{1}\mu_{N_f-N_c-1},\mu_{N_c+2}\mu_{N_f}$ & $ \nu$ \\ \hline
              $(B\otimes B)_{\nu^2}=S\sum_{i=0}^{\lfloor (N_c-1)/2 \rfloor}\text{tr}S^{2i+1}* \wedge^{N_c-1-2i}M \equiv 0$  & $2N_c+2$ & $\mu_{N_c-1}\mu_{N_c+3},\mu_{N_c-3}\mu_{N_c+5},\dots$ & $\nu^2$ \\ \hline
              
              $(B\otimes B)_1=\sum_{i=0}^{\lfloor N_c/2\rfloor}\text{tr}S^{2i+1}*\wedge^{N_c-2i}M \equiv 0$ & $2N_c+2$ & $\mu_{N_c}\mu_{N_c+2},\mu_{N_c-2}\mu_{N_c+4},\dots$ & 1 \\ \hline
            \end{tabular}
            \caption{Chiral ring constraints of strongly coupled $SU(N_c)_{1/2}+(2N_c+1)F$. An example is worked out in \S\ref{sec:region31/2}. For the properties of Casimirs of $S$, the last two constraints are identically zero.}
            \label{tab:region3phase2rel}
        \end{table}

\item Phase III exists for any $k>1$ inside region III. This phase is relatively simple because it only has enhancement to $SU(N_f+1)\times U(1)$. The chiral ring generators are reported in Tab.~\ref{tab:gensreg3fas3} and the relations in Tab.~\ref{tab:region3fase3}.
   \renewcommand{\arraystretch}{1.3}
        \begin{table}[h]
            \centering
            \begin{tabular}{|c|c|c|c|} \hline 
                & R-charge  & $SU(N_f+1)$ & $U(1)$ \\ \hline \hline
               $M_i^j$  & 2 &  $\mu_1\mu_{N_f}$ & 0 \\ \hline
               $S$ & 2 & 1 & 0 \\ \hline
               $I_{i_1 \dots i_b}$ & $N_f+|k|$ & $\mu_{b}$ & $+1$ \\ \hline
               $\tilde{I}_{i_1 \dots i_{N_f+1-b}}$ & $N_f+|k|$ & $\mu_{N_f+1-b}$ & $-1$ \\ \hline
            \end{tabular}
            \caption{Chiral ring generators of Phase III, region III. The label $b$ represents $b=|k|-1$, for ease of notation.}
            \label{tab:gensreg3fas3}
        \end{table}
 \renewcommand{\arraystretch}{1.8}
 \begin{table}[h]
            \centering
            \begin{tabular}{|c|c|c|c|} \hline 
                & R-charge  & $SU(N_f+1)$ & $U(1)$ \\ \hline \hline
               $M^2=S^2$  & 4 &  $\mu_1\mu_{N_f}+1$ & 0 \\ \hline
               $I\times M=SI$ & $N_f+|k|+2$ & $\mu_b,\mu_{b+1}\mu_{N_f},\mu_{1}\mu_{N_f-b}$ & 1 \\ \hline
               \makecell{$I\otimes \tilde{I}=S^{N_f+|k|}\exp_{b}(M/S)$} & $2N_f+2|k|$ & \makecell{$\sum_{k=0}^{b}\mu_{b-k}\mu_{N_f-b+1+k}$} & 0 \\ \hline
            \end{tabular}
            \caption{Chiral ring constraints of $SU(N_c)_k+(2N_c+2-2|k|)$ at strong coupling. The label $b$ represents $b=|k|-1$. An example is worked out in \S\ref{sec:region3greater}.}
            \label{tab:region3fase3}
        \end{table}

\end{itemize}

\vspace{-0.3cm}
\subsection{Region IV: $|k|=N_c-N_f/2+2$}
\begin{figure}[h]
    \centering
    \includegraphics[width=0.65\textwidth]{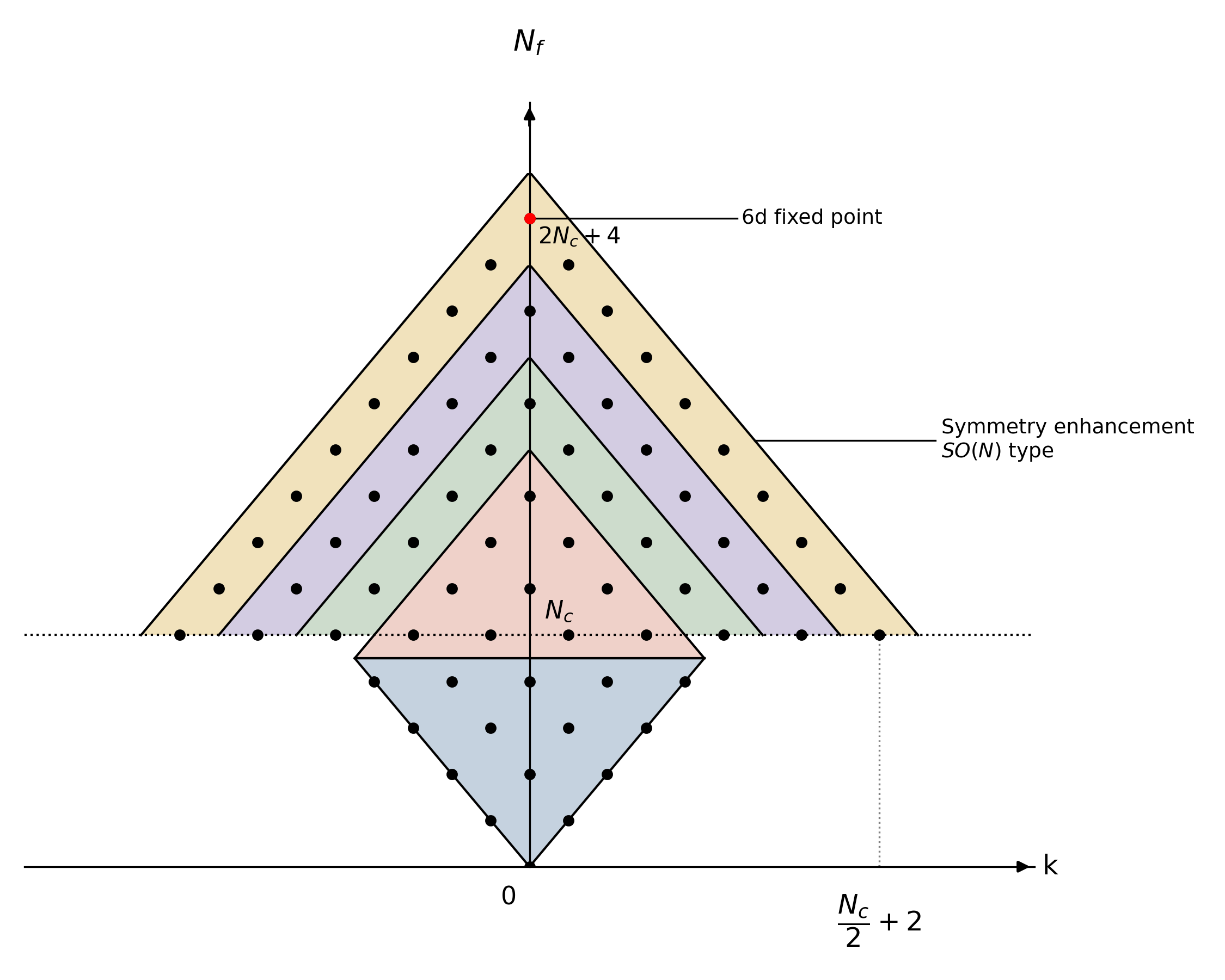}
    \caption{Phase diagram of strongly coupled $SU(N_c)_{k}+N_f$, including all regions. Only the phases with generators of R-charge $R>2$ are shown. The reference value of $N_c$ in the diagram is $N_c=5$.}
    \label{fig:region4}
\end{figure}

Region IV is defined for $|k|=N_c+2-N_f/2$; the various phases therein are identified by different patterns of symmetry enhancement. The classical symmetry enhances to either $SO(2N_f)$, $SO(2N_f)\times U(1)$, $SO(2N_f)\times SU(2)$ or $SO(2N_f+2)$.
These phases of this region are in 1-to-1 correspondence with the chiral ring of $Sp(N_c)$ gauge theories and its pattern of symmetry enhancement, as it's clear by comparing Tab.~16, 17, 18, 19, 20 of \cite{cabrera2019tropical} and Tab.~8 of \cite{Hanany:2025ctg} (which starts from the original results of \cite{bourget2020magnetic}). We thus refer to \cite{Hanany:2025jwo} for the analysis of generators and relations. The complete phase diagram is reported in Fig.~\ref{fig:region4}.
The value of $(N_f,k)=(2N_c+4,0)$ corresponds to a 6d fixed point, instead of a 5d, a phenomenon which is already well known from $SU(2)+8F$. The distinction between theories coming from compactification from 6d $(0,2)$ and genuine 5d theory has been elaborated upon in \cite{Jefferson:2017ahm}.

\subsection{General pattern}\label{sec:generalpattern}
The form of the chiral ring relations for SU theories obeys a common structure in all phases, which is the one in Tab.~\ref{tab:condensedconstraintsphaseI}, however it is particularly simple to read for those phases without symmetry enhancement. In particular, it was found in \cite{Hanany:2025jwo} that the instanton--anti-instanton equation for $SO(2N_f)\times U(1)$ global symmetry is:
\begin{equation}\label{eq:SOspinors}
    I\otimes\tilde{I}=S^{h^{\vee}}\exp\bigg(\dfrac{M}{S}\bigg)=S^{h^{\vee}}\sum_{i=1}^{N_f}\wedge^{i}\bigg(\dfrac{M}{S}\bigg)
\end{equation}
with the instanton in the spinor representation. For $SU(N_c)$ theories instead we find a correction due to the Chern-Simons level:
\begin{equation}\label{eq:SUspinors}
    I\otimes \tilde{I}=S^{h^{\vee}}\sum_{i=0}^{N_f/2-|k|}\wedge^{i}\bigg(\dfrac{M}{S}\bigg) \ .
\end{equation}
This implies that while in Eq.~\ref{eq:SOspinors} the representations appearing on the r.h.s. of the Fierz identity span all allowed values, in Eq.~\ref{eq:SUspinors} they are bounded by $|k|$.\\

An additional effect of the Chern-Simons level is symmetry enhancement. Indeed, we have observed that the pattern of enhancement is strongly dependent on $k$, inside each region. Furthermore, we observe that smaller values of $k$ correspond, on average, to larger enhancement of the symmetry as reported in Tab.~\ref{tab:symmetryandCS}.
\begin{table}[h]
    \centering
    \begin{tabular}{|c||c|c|c|} \hline 
        Region &  $k=0$ & $k=1/2$ & $k =1$\\ \hline \hline
        II & $ SU(2)\times SU(2)$ & $  SU(2)\times U(1)$ & $ SU(2)\times U(1)$ \\ \hline 
        III & $SU(N_f+2)$ & $SU(N_f+1)\times SU(2)$ & $SU(N_f+1)\times U(1)$ \\ \hline
        IV & 6d fixed point & $SO(2N_f+2)$ & $SO(2N_f)\times SU(2)$ \\ \hline
    \end{tabular}
    \caption{Dependence of the enhancement of the classical symmetry $SU(N_f)\times U(1)_I \times U(1)_B$ upon the Chern-Simons level.}
    \label{tab:symmetryandCS}
\end{table}

Moreover, instantons and baryons of $SU(N_f)$ share a kind of purity condition, as in the $SO(2N_f)$ case. Indeed, $SO(2N_f)$ spinors were found to be pure spinors, in the sense that:
\begin{equation}
    \text{Sym}^2I= A_{(N_f)} \ ,
\end{equation}
meaning that their bilinear only contains a dual (or self-dual) form, while all lower forms are constrained to be vanishing. The same condition on the symmetric product of instantons and baryons are found, as reported in Tab.~\ref{tab:condensedconstraintsphaseI} and Tab.~\ref{tab:constraintsphaseII}.
This condition allowed to interpret the chiral ring relations of $Sp(N_c)$ theories in \cite{Hanany:2026eyi}, using the well-developed techniques of pure spinors. Indeed, it was possible to find that, out of all chiral ring relations, only the purity condition and the instanton bilinear equations were independent. Furthermore, it was possible to recover the stratification of the Higgs branch, without referring to \emph{quiver subtraction} conjectures.
Similar techniques are less-developed for $SU(N_f)$ ``pure" instantons and baryons, hence it is not clear how to interpret geometrically those chiral ring relations.\\

As a final comment, the chiral ring of $Sp(N_c)$ was found \cite{Hanany:2025jwo} to be corrected even at small coupling, by putting the instanton and anti-instanton to zero in the chiral ring relations. The same feature can be easily seen for $SU(N_c)$ theories; this does not imply that the Higgs branch metric is quantum corrected at small coupling, as indeed the presence of a Lagrangian description allows to make use of non-renormalization arguments. These finite-coupling chiral ring corrections are yet to be found using Lagrangian methods.
 
\section{Chiral ring relations of SO theories}\label{sec:SO}
$SO(N_c)$ theories at finite coupling admit mesons and baryons. At infinte coupling \cite{Akhond:2024nyr} however, the UV fixed point exists only for $N_f\leq N_c-3$, hence baryonic phases are cut off. Similarly to the $SU(N_c)$ case, there is a subregion of the $(N_f,N_c)$ lattice characterized by no symmetry enhancement, and there are two discrete choices ($N_f=N_c-3$ and $N_f=N_c-4$) with patterns of symmetry enhancement.
 
\subsection{Low flavours: $N_f<N_c-4$}
In this phase, the global symmetry is $Sp(N_f)\times U(1)_I$, the same as in the IR limit. The generators are reported in Tab.~\ref{tab:SOgensNOenh}

\begin{table}[h]
    \centering
    \begin{tabular}{|c|c|c|c|} \hline
        Generators & R-charge & $Sp(N_f)$ & $U(1)_I$ \\ \hline
        $M_{(ij)}$ & 2 & $\mu_1^2$ & 0 \\ \hline
        $S$ & 2 & 1 & 0 \\ \hline
        \makecell{$I_{[i_1\dots i_{N_f}]}$\\$\tilde{I}_{[i_1\dots i_{N_f}]}$} & $N_c-2$ & $\mu_{N_f}$ & $\pm 1$ \\ \hline
    \end{tabular}
    \caption{Generators of the chiral ring at infinite coupling of $SO(k)$ with $N_f<N_c-4$ fundamentals. The instanton and anti-instanton transform in irreducible representations, hence they are defined such that any contraction with the symplectic invariant $\Omega_{ij}$ vanishes.}
    \label{tab:SOgensNOenh}
\end{table}
Let us now study the constraints among these using naturalness arguments. We will then check these hypothesis for a few examples.
\begin{itemize}
    \item 
\textbf{Involving $M_{(ij)}$ and $S$.} There are two relations involving mesons and gaugino bilinear:
\renewcommand{\arraystretch}{1.5}
\begin{table}[H]
    \centering
    \begin{tabular}{|c|c|c|c|}\hline
$M_{(ij)}M_{(kl)}\Omega^{[ik]}\Omega^{[jl]}=S^2$  & 4 & 1 & 0  \\ \hline
  $M_{(ij)}M_{(kl)}\Omega^{[jk]}-\Omega_{il}(M\cdot M)/{2N_f}=0$  & 4 & $\mu_2$ & 0 \\  \hline
    \end{tabular}
    \caption{Mesons and gaugino relations. The two relations can be presented as a single equation in a reducible representation.}
    \label{tab:SO,mANDs}
\end{table}
Indeed the product of the mesons, from a representation theoretic point of view, is:
\[
\text{Sym}^2\mu_1^2= \mu_1^4+\mu_2^2+\mu_2+1
\]
with $\mu_1^4$ being the symmetrization of the four indices, $1$ being the contraction of all indices, $\mu_2$ being the contraction of one index on each meson using the Omega tensor and $\mu_2^2$ being the symmetrization of the two pairs of indices.

\item \textbf{Relations involving $S$ and $I,\tilde{I}$, at lower R-charge.}
The tensor product of the meson and instanton representations is:
\[
\mu_1^2 \otimes \mu_{N_f}=\mu_1^2\mu_{N_f}+\mu_{N_f}+\mu_1\mu_{N_f-1}
\]
where the first contribution comes from the full expression, the last two from the contraction of an index of the mesons with one of the instantons. Hence we can write:
\begin{table}[H]
    \centering
    \begin{tabular}{|c|c|c|c|} \hline
      $SI_{[ki_2\dots i_{N_f}]}+M_{kl}I_{[i_1\dots i_{N_f}]}\Omega^{li_1}=0$  & $N_c-2$ & $\mu_{N_f}+\mu_1\mu_{N_f-1}$ & $\pm 1$  \\ \hline
    \end{tabular}
    \caption{Chiral ring constraints involving mesons, gaugino bilinear and instantons, for strongly coupled $SO(N_c)$ theories.}
    \label{tab:SO,m,SandI}
\end{table}

\item \textbf{Relations involving separately $I$ and $\tilde{I}$.}
These must transform in the:
\[
\text{Sym}^2\mu_{N_f}=\mu_{N_f}^2+\mu_{N_f-2}^2+\mu_{N_f-4}^2+\dots
\]
we thus ask, following the observed pattern of $SU$ and $Sp$ gauge theories, that all representations vanish except $\mu_{N_f}^2$.

\item \textbf{Relations involving $M,S,I$ and $\tilde{I}$ at higher R-charge.}
If we consider bilinears in the instantons, they transform in:
\[
\mu_{N_f}\otimes \mu_{N_f}=\sum_{i=1}^{N_f}\mu_{i}^2 \ .
\]
This product is corrected by appropriate combinations of $M$ and $S$. In particular, let us notice that the $\mu_i^2$ representations can be found by odd antisymmetrizations of the mesons.
The constraint will thus be:
\begin{equation}
   \boxed{ I \otimes \tilde{I}= \sum_{i=0,\text{odd}}^{N_f}S^{N_c-2-i}\wedge^{i}\bigg(\dfrac{M}{S}\bigg)\equiv S^{N_c-2}\exp_{N_f}^{\text{odd}}\bigg(\dfrac{M}{S}\bigg)} \ ,
\end{equation}
\end{itemize}
where we defined an odd-elements anti-symmetrization which also includes the identity.

\subsubsection*{Summary of constraints for $N_f<k-4$}
The expected constraints, for the cases without symmetry enhancement, can be condensed in:
\begin{table}[H]
     \centering
     \begin{tabular}{|c|c|c|c|} \hline
     Constraints & R-charge & $Sp(N_f)$ & $U(1)$ \\ \hline \hline
     $M^2=S^2\Omega$  & 4 & $\mu_2+1$ & 0  \\ \hline
           $SI+M \times I=0$  & $N_c$ & $\mu_{N_f}+\mu_1\mu_{N_f-1}$ & $\pm 1$  \\ \hline
           $I\tilde{I}=S^{N_c-2}\exp_{N_f}^{\text{odd}} M/S$ & $2N_c-4$ & $\sum_{i=0}^{N_f}\mu_i^2$ & 0 \\ \hline
          $(\text{Sym}I)_{\mu_{N_f}}=(II)_{\mu_{N_f}^2}$ & $2N_c-4$ & $\sum_{k=1}\mu_{N_f-2k}^2$ & $\pm 2$ \\ \hline
     \end{tabular}
     \caption{Chiral ring constraints for $SO(N_c)$ with $N_f$ flavours, with $N_f < N_c-4$. The $\times$ symbol represents two different operations: a contraction with the symplectic invariant $\Omega$ and with the $\delta$ with upper and lower indices.}
     \label{tab:SOgeneratorsNosymmetryenhancement}
 \end{table}  

\subsubsection*{Consistency checks}
\begin{itemize}
    \item $SO(6)+1F$, (which is the same as $SU(4)$ with 1 antisymmetric), has the following generators:
    \renewcommand{\arraystretch}{1.5}
    \begin{table}[H]
        \centering
        \begin{tabular}{|c|c|c|c|} \hline
          Generators   & R-charge &  $Sp(1)$ & $U(1)$ \\ \hline
            $M_{(ij)}$ & 2 & $\mu^2$ & 0 \\ \hline
            $S$ & 2 & 1 & 0 \\ \hline
            $I_{i},\tilde{I}_{i}$ & 4 & $\mu$ & $\pm 1$ \\ \hline
        \end{tabular}
        \caption{Generators of strongly coupled $SO(6)+1F$.}
        \label{tab:SO(6)+1Fgen}
    \end{table}

and constraints:
  \begin{table}[H]
        \centering
        \begin{tabular}{|c|c|c|c|} \hline
            Constraints & R-charge & $Sp(1)$ & $U(1)$ \\ \hline \hline
           $M^2=S^2$ & 4 & $1$ & 0 \\ \hline
           $MI=SI$ & 6 & $\mu$ & $\pm 1$ \\ \hline
           $(I\tilde{I})_1=S^4$ & 8 & 1 & 0 \\ \hline
           $(I\tilde{I})_{\mu^2}=SM_{ij}$ & 8 & $\mu^2$ & 0 \\ \hline
        \end{tabular}
        \caption{Chiral ring relations of strongly coupled $SO(6)+1F$.}
        \label{tab:SO(6)+1Frel}
    \end{table}

\item SO(7); The pure theory has just $A_4$ as Higgs branch. The theory with 1 flavour is the same as the previous case, except the R-charge of the instantons is 5. The theoy with two flavours has the following generators:
  \begin{table}[H]
        \centering
        \begin{tabular}{|c|c|c|c|} \hline
          Generators   & R-charge &  $Sp(2)$ & $U(1)$ \\ \hline  \hline
            $M_{(ij)}$ & 2 & $\mu_1^2$ & 0 \\ \hline
            $S$ & 2 & 1 & 0 \\ \hline
            $I_{[ij]},\tilde{I}_{[ij]}$ & 5 & $\mu_2$ & $\pm 1$ \\ \hline
        \end{tabular}
        \caption{Chiral ring generators of $SO(7)+2F$ at strong coupling.}
        \label{tab:SO(7)+2gen}
    \end{table}
and constraints:
\begin{table}[H]
    \centering
    \begin{tabular}{|c|c|c|c|} \hline 
       Constraints  & R-charge & $Sp(2)$ & $U(1)$ \\ \hline \hline
       $M^2_{[ij]}=S^2\Omega_{ij}$  & 4 & $\mu_2+1$ & 0 \\ \hline
       $MI=SI$ & 7 & $\mu_2+\mu_1^2$ & $\pm 1$ \\ \hline
        $(I\tilde{I})_{1}=S^5$ & 10 & 1 & 0 \\ \hline
         $(I\tilde{I})_{\mu_1^2}=S^4M$ & 10 & $\mu_1^2$ & 0 \\ \hline
          $(I\tilde{I})_{\mu_2^2}=S^2\wedge^3M$ & 10 & $\mu_2^2$ & 0 \\ \hline
       $I \cdot I=0$ & 10 & 1 & $\pm 2$ \\ \hline
    \end{tabular}
    \caption{Chiral ring relations of $SO(7)+2F$ at strong coupling.}
    \label{tab:SO(7)+2rel}
\end{table}
A further example with $SO(8)$ gauge group can be found in \S\ref{par:SO(8)}. 
\end{itemize}

\subsection{Symmetry enhancement}
$SO(2N_c)$ theories undergo partial symmetry enhancement for $N_f=N_c-4$ and maximal symmetry enhancement (of the type $Sp(N_f)\times U(1)_I\to Sp(N_f+1)$) for $N_f=N_c-3$. We will be rather schematic in this section, since the logic is the same as in the rest of the paper.
\paragraph{Partial symmetry enhancement: $N_f=N_c-4$.}
The chiral ring generators are:
\begin{table}[H]
        \centering
        \begin{tabular}{|c|c|c|c|} \hline
          Generators   & R-charge &  $Sp(N_f)$ & $SU(2)$ \\ \hline \hline
            $M_{(ij)}$ & 2 & $\mu_1^2$ & 1 \\ \hline
            $S^{(ab)}$ & 2 & 1 & $\nu^2$ \\ \hline
            $I_{[i_1\dots i_{N_f}]}^{(ab)}$ & $N_c-2$ & $\mu_{N_f}$ & $\nu^2$ \\ \hline  
        \end{tabular}
        \caption{Chiral ring generators for $SO(N_c)+(N_c-4)F$ at strong coupling. The instantons are implicitly assumed to be traceless.}
        \label{tab:SOgenPartial}
\end{table}
and the relations are:
\begin{table}[H]
        \centering
        \begin{tabular}{|c|c|c|c|} \hline
          Generators   & R-charge &  $Sp(N_f)$ & $SU(2)$ \\ \hline \hline
            $M^2=S^2$ & 4 & $\mu_2+1$ & 1 \\ \hline
            $M \times I=SI$ & $N_c$ & $\mu_{N_f}+\mu_{N_f-1}\mu_1$ & $\nu^2$ \\ \hline
                 \makecell{$I\otimes I=\text{tr}S^{N_c-2}\text{Sym}^2S$\\ if $N_c$ even} & $2N_c-4$ & 1 & $1+\nu^4$ \\ \hline
            \makecell{$I\otimes I=(\text{Sym}^2S)\sum_i\text{tr}S^{N_c-3-2i}\wedge^{2i+1}M$\\ if $N_c$ odd} & $2N_c-4$ & $\sum_i\mu_i^2$ & $1+\nu^4$ \\ \hline
        \end{tabular}
        \caption{Chiral ring generators for $SO(N_c)+(N_c-4)F$ at strong coupling. An example is worked out in \S\ref{par:exSO}. The distinction between even and odd flavours descends from odd Casimirs of $S$ being vanishing.}
        \label{tab:SOrelPartial}
\end{table}

\paragraph{Maximal symmetry enhancement: $N_f=N_c-3$.}
The chiral ring generators are reported in Tab.~\ref{tab:Somaximalgen} and the constraints in Tab.~\ref{tab:Somaximalrel}.
\renewcommand{\arraystretch}{1.5}
\begin{table}[h]
        \centering
        \begin{tabular}{|c|c|c|} \hline
          Generators   & R-charge &  $Sp(N_f+1)$ \\ \hline
            $M_{(ij)}$ & 2 & $\mu_1^2$  \\ \hline
            $I_{[i_1\dots i_{N_f+1}]}$ & $N_c-2$ & $\mu_{N_f+1}$  \\ \hline
        \end{tabular}
        \caption{Chiral ring generators of strongly coupled $SO(N_c)+(N_c-3)$ flavours. The instantons are implicitly assumed to be traceless.}
        \label{tab:Somaximalgen}
    \end{table}
\begin{table}[h]
        \centering
        \begin{tabular}{|c|c|c|} \hline
          Generators   & R-charge &  $Sp(N_f+1)$ \\ \hline
            $M^2=0$ & 4 & $\mu_2$  \\ \hline
            $M \times I=0$ & $N_c$ & $\mu_{N_f+1}+\mu_1\mu_{N_f}$  \\ \hline
            \makecell{$I\otimes I=\text{tr}M^{N_c-2}$\\ if $N_c$ even} & $2N_c-4$ & 1 \\ \hline
             \makecell{$I\otimes I=\sum_i\text{tr}M^{N_c-3-2i}\wedge^{2i+1}M$\\ if $N_c$ odd} & $2N_c-4$ & $\sum_{i}\mu_i^2$ \\ \hline
        \end{tabular}
        \caption{Chiral ring relations of strongly coupled $SO(2N_c)+(N_c-3)$ flavours.}
        \label{tab:Somaximalrel}
    \end{table}

\section{Lagrangian submanifold}\label{sec:Lagrangian}
The structure of the holomorphic functions on the Higgs branch $\mathcal{H}$ allows to identify Lagrangian submanifolds of the variety itself. These submanifolds are the image $\mathcal{B}$ of holomorphic functions $f_i$, such that:
\begin{equation}
    \begin{cases}
        \text{dim}\mathcal{B} & =\dfrac{1}{2}\text{dim}\mathcal{H} \ ,\\
        \{f_i,f_j\} & =0 \ ,
    \end{cases}
\end{equation}
where the brackets represent the Poisson structure. Lagrangian submanifolds are interesting from the point of view of integrability, deformation quantization and potentially a twistor formulation of the Higgs branch, and will be the subject of future work as well.\\

The way these Lagrangian submanifolds are identified is the following: we consider the \emph{highest weight generating} function (HWG) \cite{hanany2014highest} of the chiral ring. In many cases, the HWG can be reformulated in the following form:
\begin{equation}
    \text{HWG}=\bigg(\text{PE}[\mathcal{I}^+]+\text{PE}[\mathcal{I}^-]-1\bigg)\cdot (\text{dressing)}
\end{equation}
where $\mathcal{I}^+$ and $\mathcal{I}^-$ are (the highest weight representations of) the holomorphic functions spanning two orthogonal cones, with disjoint charges with respect to $U(1)$. The dressing part of the $\text{HWG}$ is charge-independent and corresponds to the HWG of the theory at finite coupling.
The space of functions generated by $\mathcal{I}^+$ (or equivalently $\mathcal{I}^-$):
\[
\text{PE}[\mathcal{I}^+] \ ,
\]
maps to a Lagrangian submanifold of the Higgs branch: it is half-dimensional and the operator represented by $\mathcal{I}^+$ commutes with itself. In the original paper \cite{Hanany:2025ctg}, to which we refer for more details, $\mathcal{I}^+$ was called a \emph{bare instanton}, since it was charged under $U(1)_I$, and the space spanned by the associated holomorphic function $I^+$ was called \emph{bare instanton space}. We suggest in general to call such holomorphic functions \emph{bare action variables}, since they may be charged under different $U(1)$'s.
The bare action variables are the highest weight of the representation of the holomorphic functions; for example, $\mathcal{I}^+$ is the highest weight representation of the representation in which the instanton operator $I_{i_1 \dots i_n}^+$ of positive charge transforms.
\\

In the rest of this section we will apply this same technique to $SU$ and $SO$ theories in various phases; it will not always be possible to find such dressed action variable construction, as in many cases with enhanced symmetry the structure of the action variables is more elaborate. For example, as reported in \cite{cremonesi2017instanton, Hanany:2025jwo}, for $Sp(N_c)$ with $N_f=2k+4$ flavours, the Lagrangian submanifold is the pure spinor cone endowed with an additional orthogonal line. For $N_f=2k+5$ flavours, instead, the Lagrangian submanifold is the pure spinor cone enlarged with a quadric. The orthogonality of these spaces can be read from the chiral ring relations in \cite{cremonesi2017instanton} by setting $M$ and $S$ to zero.

\subsection{Dressed action variables for SU theories}
\begin{itemize}
    \item Region 1; let us report in Tab.~\ref{tab:region1HWG} the HWGs for this region, as found in \cite{Zhong:2022pmg}.
\renewcommand{\arraystretch}{2.2}
\begin{table}[h]
    \centering
    \begin{tabular}{|c|c|} \hline
       Branch  &  $\text{PL}[\text{HWG}]$ \\ \hline \hline
       Instantonic  &  $\sum_{i=1}^{N_f/2-|k|}\mu_i\mu_{N_f-i}t^{2i}+t^2+q\mu_{N_f/2-|k|}t^{N_c}+q^{-1}\mu_{N_f/2+|k|}t^{N_c}-\mu_{N_f/2-|k|}\mu_{N_f/2+|k|}t^{2N_c}$ \\ \hline
       Baryonic & $\sum_{i=1}^{N_f-N_c}\mu_i\mu_{N_f-i}t^{2i}+t^2+q\mu_{N_c}t^{N_c}+q^{-1}\mu_{N_f-N_c}t^{N_c}-\mu_{N_c}\mu_{N_f-N_c}t^{2N_c}$ \\ \hline 
       \makecell{Mesonic \\ ($N_f$ even)} & $\sum_{i=1}^{N_f/2}\mu_i\mu_{N_f-i}t^{2i}$ \\ \hline
       \makecell{Mesonic \\ ($N_f$ odd)}  & $\sum_{i=1}^{(N_f-1)/2}\mu_i\mu_{N_f-i}t^{2i}$\\ \hline \hline
       \makecell{I $\cap$ II \\ II $\cap$ III \\ I $\cap$ II $\cap$ III } & $\sum_{i=1}^{N_f-N_c}\mu_i\mu_{N_f-i}t^{2i}$\\ \hline
              I $\cap$ III & $\sum_{i=1}^{N_f/2-|k|}\mu_i\mu_{N_f-i}t^{2i}$\\ \hline
    \end{tabular}
    \caption{HWGs for region I, with $N_c-N_f/2>|k|$. Phase I satisfies $N_f/2 \geq |k|$, phase II is present only for $N_f \geq N_c$. The quantity $N_f/2\pm|k|$ is always an integer.}
    \label{tab:region1HWG}
\end{table}
    \begin{itemize}
    \item Phase I can be build as moduli space of dressed instantons; the bare instanton at charge $i$ has the following highest weight representation:
      \[
    \mathcal{I}_i=\mu_{N_f/2-|k|}^it^{iN_c} \ ,
    \]
    and
    \[
     \mathcal{I}_{-i}=\mu_{|k|+N_f/2}^it^{iN_c} \ ,
    \]
    while the dressing factor is:
    \[
    \text{HWG}_0=\text{PE}\bigg[\sum_{i=1}^{N_f/2-|k|}\mu_i\mu_{N_f-i}t^{2i} \bigg]\ .
    \]
    The space of bare instantons of positive (or negative) charge is indeed a Gorenstein singularity of complex dimension:
    \[
\text{dim}_{\mathbb{C}}=\dfrac{N_f^2}{4}-k^2+1 \ .
    \]
    The space of bare instantons is half-dimensional with respect to phase I; furthermore, instanton operators Poisson-commute with themselves. 
    \item Branch II can be built as the moduli space of dressed baryon operators; the bare baryons at charge $j$ has the following highest weight representation:
        \[
    \mathcal{B}_j=\mu_{N_f-N_c}^jt^{jN_c}
    \]
    and
    \[
     \mathcal{B}_{-j}=\mu_{N_c}^jt^{jN_c}
    \]
    while the dressing factor is:
    \[
    \text{HWG}_0=\text{PE}\bigg[\sum_{i=1}^{N_c-N_f}\mu_i\mu_{N_f-i}t^{2i}\bigg]
    \]
    The bare baryons space is a Gorenstein singularity, of complex dimension $N_cN_f-N_c^2+1$. 
\end{itemize}

\item Region II. The corresponding HWG are reported in Tab.~\ref{tab:region2HWG}.
\renewcommand{\arraystretch}{2.2}
\begin{table}[h]
    \centering
    \begin{tabular}{|c|c|c|} \hline
       Branch  &  Symmetry & $\text{PL}[\text{HWG}]$ \\ \hline \hline
       \makecell{I \\ $k=0$}  & $SU(N_f)\times SU(2)\times SU(2)$  & \makecell{$\sum_{i=1}^{N_f/2}\mu_i\mu_{N_f-i}t^{2i}+t^4+\nu_1t^2+\nu_2t^2+\nu_1\nu_2\mu_{N_f/2}t^{N_c}$\\$+\nu_1\nu_2\mu_{N_f/2}t^{N_c+2}-\nu_1^2\nu_2^2\mu_{N_f/2}^2t^{2N_c+4}$} \\ \hline
       \makecell{I\\ $k \geq 1/2$} & $SU(N_f)\times SU(2)\times U(1)$ & \makecell{$\sum_{i=1}^{N_f/2-|k|}\mu_i\mu_{N_f-i}t^{2i}+t^2+\nu^2t^2+\nu\mu_{N_f/2-|k|}q t^{N_c}$\\$+\nu\mu_{N_f/2+|k|}q^{-1}t^{N_c}-\nu^2\mu_{N_f/2-|k|}\mu_{N_f+|k|}t^{2N_c}$} \\ \hline 
       \makecell{III \\  $k=1$, ($N_f$ even)} & $SU(N_f)$ & $\sum_{i=1}^{N_f/2}\mu_i\mu_{N_f-i}t^{2i}$ \\ \hline
        \makecell{III \\  $k>1$, ($N_f$ even)} & $SU(N_f)\times SU(2)$ &$\sum_{i=1}^{N_f/2}\mu_i\mu_{N_f-i}t^{2i}+\nu^2t^2$ \\ \hline
       \makecell{III \\  $k=1$, ($N_f$ odd)}  & $SU(N_f)$ & $\sum_{i=1}^{(N_f-1)/2}\mu_i\mu_{N_f-i}t^{2i}$\\ \hline 
       \makecell{III \\  $k>1$, ($N_f$ odd)}  & $SU(N_f)\times SU(2)$ & $\sum_{i=1}^{(N_f-1)/2}\mu_i\mu_{N_f-i}t^{2i}+\nu^2t^2$\\ \hline \hline
    \makecell{I $\cap$ III\\ $k=1$} & $SU(N_f)$ & $\sum_{i=1}^{(N_f-2)/2-|k|}\mu_i\mu_{N_f-i}t^{2i}$\\ \hline
    \makecell{I $\cap$ III\\ $k>1$} & $SU(N_f)\times SU(2)$ & $\sum_{i=1}^{N_f/2-|k|}\mu_i\mu_{N_f-i}t^{2i}+\nu^2 t^2$\\ \hline 
    \end{tabular}
    \caption{HWGs for region II, with $N_c-N_f/2=|k|$. Phase I satisfies $N_f/2 \geq |k|$, phase III is present for all $N_f \geq 1$ and $|k|\geq 1/2$. The quantity $N_f/2\pm|k|$ is always an integer.}
    \label{tab:region2HWG}
\end{table}
\begin{itemize}
   \item Phase I, $k\geq 1/2$. The bare action $\mathcal{I}^+,\mathcal{I}^-$ variables at $U(1)$ charge $i$ have highest weight representations:
   \begin{equation}\label{eq:dressedregion2}
 \begin{array}{ll}
      (\mathcal{I}^+)^i  & =  \nu^i \mu_{N_f/2+|k|}^i t^{iN_c}\\
       (\mathcal{I}^-)^{-i}  & =  \nu^i \mu_{N_f/2-|k|}^i t^{iN_c}\\
       \end{array}
   \end{equation}
  with the dressing factor being:
  \begin{equation}
\text{HWG}_0=\text{PE}\bigg[t^4+\nu^2t^2+\sum_{i=1}^{N_f/2-|k|}\mu_i\mu_{N_f-i}t^{2i}\bigg]
  \end{equation}
\end{itemize}
It is not clear instead how to find a polarization for Phase I $k=0$. Nor is it obvious that the choice of action variables for Phase I $k \geq 1/2$ is unique, since in Eq.~\ref{eq:dressedregion2} we have considered the $U(1)$ grading, instead of the $SU(2)$.

\item Region 3; the HWG's are reported in Tab.~\ref{tab:region3HWG}.
\renewcommand{\arraystretch}{2.4}
\begin{table}[h]
    \centering
    \begin{tabular}{|c|c|c|} \hline
       Branch  &  Symmetry & $\text{PL}[\text{HWG}]$ \\ \hline \hline
       \makecell{I \\ $k=0$}  & $SU(N_f+2)$  & \makecell{$\sum_{i=1}^{N_f/2}\mu_i\mu_{N_f+2-i}t^{2i}+t^4+\mu_{N_f/2+1}t^{N_c}+\mu_{N_f/2+1}t^{N_c+2}$} \\ \hline
       \makecell{I\\ $k = 1/2$} & $SU(N_f+1)\times SU(2)$ & \makecell{$\sum_{i=1}^{(N_f+1)/2}\mu_i\mu_{N_f+1-i}t^{2i}+t^4+\nu^2t^2+\nu\mu_{(N_f+1)/2} t^{N_c+1}$\\$+\nu\mu_{(N_f+1)/2}t^{N_c+3}-\nu^2\mu_{(N_f+1)/2}^2t^{2N_c+4}$} \\ \hline 
         \makecell{I\\ $k \geq 1$} & $SU(N_f+1)\times U(1)$ & \makecell{$\sum_{i=1}^{N_f/2-|k|+1}\mu_i\mu_{N_f+1-i}t^{2i}+t^2+q\mu_{N_f-|k|+1} t^{N_f+|k|}$\\$+q^{-1}\mu_{|k|-1}t^{N_f+|k|}-\mu_{|k|}\mu_{N_f+1-|k|}t^{2N_f+2|k|}$} \\ \hline 
       \makecell{III \\  $k=3/2$, ($N_f$ even)} & $SU(N_f+1)$ & $\sum_{i=1}^{N_f/2}\mu_i\mu_{N_f+1-i}t^{2i}$ \\ \hline
       \makecell{III \\  $k=3/2$, ($N_f$ odd)}  & $SU(N_f+1)$ & $\sum_{i=1}^{(N_f+1)/2}\mu_i\mu_{N_f+1-i}t^{2i}$\\ \hline \hline
    \makecell{I $\cap$ III\\ $k>3/2$} & $SU(N_f+1)$ & $\sum_{i=1}^{N_f/2-|k|+1}\mu_i\mu_{N_f+1-i}t^{2i}$\\ \hline
    \end{tabular}
    \caption{HWGs for region III, with $N_c-N_f/2+1=|k|$. Phase I satisfies $N_f/2 \geq |k|-1$, phase III is present for all $N_f \geq 1$ and $|k|\geq 1/2$. The quantity $N_f/2\pm|k|$ is always an integer.}
    \label{tab:region3HWG}
\end{table}
For phase I with $k\geq 1/2$ the situation is completely analogous to the one described in \cite{Hanany:2025ctg}, except for the substitution of $SO(2N_f)$ with $SU(N_f+1)$, hence we refer the reader to that treatment. For the other cases, the same problems persist in finding a polarization.
\item Region 4; the HWG's are reported in Tab.~\ref{tab:region4hwg}.
\renewcommand{\arraystretch}{2.2}
\begin{table}[h]
    \centering
    \begin{tabular}{|c|c|c|} \hline
       Branch  &  Symmetry & $\text{PL}[\text{HWG}]$ \\ \hline \hline
       \makecell{I (even) \\ $k>2$}  & $SO(2N_f)$  & \makecell{$\sum_{i=1}^{N_f/2-1}\mu_{2i}t^{2i}+\mu_{N_f}^2t^{N_f}$} \\ \hline
       \makecell{II (even)\\ $k =1$} & $SO(2N_f)\times SU(2)$ & \makecell{$\sum_{i=1}^{N_f/2-1}\mu_{2i}t^{2i}+t^4+\nu^2t^2+\nu\mu_{N_f}\big( t^{N_c}+t^{N_c+2}\big)+$\\$+\mu_{N_f}^2t^{2N_c+2}-\nu^2\mu_{N_f}^2t^{2N_c+4}$} \\ \hline 
       \makecell{II (even)\\ $k >1$} & $SO(2N_f)\times U(1)$ & \makecell{$\sum_{i=1}^{N_f/2-1}\mu_{2i}t^{2i}+t^2+\mu_{N_f}(q+q^{-1})t^{N_c}+$\\$+\mu_{N_f}^2t^{N_f}-\mu_{N_f}^2t^{2N_c}$} \\ \hline 
           \makecell{II (odd)\\ $k =1/2$} & $SO(2N_f+2)$ & \makecell{$\sum_{i=1}^{(N_f-1)/2}\mu_{2i}t^{2i}+t^4+\mu_{N_f+1}\big( t^{N_c}+t^{N_c+2}\big)$} \\ \hline 
              \makecell{II (odd)\\ $k >1$} & $SO(2N_f)\times U(1)$ & \makecell{$\sum_{i=1}^{(N_f-1)/2}\mu_{2i}t^{2i}+t^2+\mu_{N_f-1}qt^{N_c}+\mu_{N_f}q^{-1}t^{N_c}+$\\$+\mu_{N_f}\mu_{N_f-1}t^{N_f-1}-\mu_{N_f}\mu_{N_f-1}t^{2N_c}$} \\ \hline 
    \end{tabular}
    \caption{HWGs for region IV, with $N_c-N_f/2+2=|k|$. Phases I and II exist for $N_f \geq 0$, with little adjustments for even/odd flavours. The quantity $N_f/2\pm|k|$ is always an integer.}
    \label{tab:region4hwg}
\end{table}
The HWGs in Tab.~\ref{tab:region4hwg} are exactly the same as those in Tab.~7 of \cite{Hanany:2025ctg}, where dressed action variables were originally studied.
\end{itemize}
\subsection{Dressed action variables for SO theories}
The HWGs are reported in Tab.~1 of \cite{Akhond:2024nyr}.
\begin{itemize}
    \item For those cases without symmetry enhancement the dressing factor is just:
    \[
\text{HWG}_0=\text{PE}\bigg[\sum_{i=1}^{N_f}\mu_i^2t^{2i}\bigg]
    \]
    which is the HWG of the (closure of the) maximal nilpotent orbit of height 2 of $C_{N_f}$. The bare instanton at charge $i$ has highest weight representation:
    \[
    \mathcal{I}_i=\mu_{N_f}^it^{i(K-2)}
    \]
    In particular, the space of dressed instantons can be checked to be a Gorenstein variety, of complex dimension:
    \[
    \text{dim}_{\mathbb{C}}=\dfrac{N_f(N_f+1)}{2}+1
    \]
\item For partial symmetry enhancement, the global symmetry $SU(2)$ can be branched into U(1), obtaining the following HWG:
\begin{equation}
\text{HWG}=\text{PL}\bigg[\sum_{i=1}^{K-4}\mu_i^2t^{2i}+(1+q^2+q^{-2})(\mu_{K-4}t^{K-2}+t^2)-\mu_{K-4}t^K-\mu_{K-4}^2t^{2K-4}\bigg]
\end{equation}
the dressing factor is:
\begin{equation}
    \text{HWG}_0=\text{PE}\bigg[\sum_{i=1}^{K-4}\mu_i^2t^{2i}-\mu_{K-4}t^K\bigg]
\end{equation}
while bare action variables at representation $[2i]$ of SU(2) are:
\begin{equation}
    \mathcal{I}^i=\mu_{K-4}^it^{i(K-2)}
\end{equation}
\item For maximal symmetry enhancement, it is not possible to find a polarization of the Higgs branch in the current understanding.
\end{itemize}

\section{Summary and outlook}
It has been found that the general pattern of chiral ring relations, for $SO,SU$ and $Sp$ gauge groups, is:
\begin{equation}\begin{cases}
    \begin{array}{ll}
       M^2=S^2  &  \\
        M \times I=SI & \\
        I\otimes \tilde{I}=S^{R}\exp_*(M/S) & \\
        \text{Sym}^2I =(II)_{\mu_{i}^2}
    \end{array}
    \end{cases}\label{eq:results}
\end{equation}
with $I$ representing the generator of R-charge $R$ and in the $\mu_i$ representation of the flavour group. The asterisk $*$ represents the restrictions due to the CS level, as discussed in \S\ref{sec:generalpattern}. This same pattern is respected, with the necessary corrections, in the phases with symmetry enhancement. While we can make sense of this pattern \emph{a posteriori}, through a \emph{naturalness} principle,  arguing that every relation with the right quantum numbers to be corrected is indeed corrected, a geometrical interpretation of these relation has been worked out only for $Sp(N_c)$ theories \cite{Hanany:2026eyi}. It would be desirable to obtain such an interpretation to the other theories as well. \\

A further novel aspect discussed in this paper concerns the baryon relations. As remarked, baryons, mesons and gaugino bilinear are classical objects described by a Lagrangian. Nevertheless, the relations in Tab.~\ref{tab:constraintsphaseII} are quite different from the one usually presented for baryons and mesons, as they also include corrections by the gaugino bilinear. It remains an open problem to reproduce the effect of the gaugino bilinear from perturbative techniques at finite couplings.\\

For what concerns the symmetry enhanced phases, it would also be desirable to extend the systematic identification of dressed action variables. However, while for regular phases the existence of a Lagrangian submanifold can be argued rather easily using the chiral ring relations and ansatz for the Poisson brackets \cite{gledhill2023poisson}, such arguments are hard to translate for enhanced phases and it is not clear whether a Lagrangian submanifold should exist at all.
In this context, it would be also interesting to understand whether the Lagrangian submanifold is unique, or whether multiple symmetry factors (like $U(1)\times SU(2)$) allow for multiple such submanifolds.\\

On a final note, while the identification of the chiral ring relations and Lagrangian submanifolds is necessary for a better understanding of 5d vacua, this is not the end of the story. We have argued that the Higgs branches are the phase spaces of classical integrable systems, and the identification of the system itself will be performed in upcoming work \cite{EVnew}. While these kinds of identification have been performed for other moduli spaces (see for example \cite{Gorsky:2000px, Chaban:2025mjz}), they usually concern Coulomb branches rather than Higgs branches. Moreover, it could be interesting to quantize the integrable system and compare it with the deformation quantization \cite{beem2017deformation} of the holomorphic functions of the chiral ring.

\vspace{1cm}
\subsection*{Acknowledgments}
 We would like to thank the Simons Center for Geometry and Physics, and the organizers of the 2026 Simons Physics Summer Workshop, where part of this work was carried out. The work of A. Hanany and E. Van den Driessche is partially supported by STFC Consolidated Grant ST/X000575/1.
\newpage
\section{Appendix: Examples}
We provide some examples for the chiral ring equations described in the paper.
\subsection*{Region 1, Phase I.}\label{par:firstexample}
To access the whole relations, let us consider the Higgs branch of $SU(4)_{0}+6$, whose Hilbert series is:
   \begin{equation}
            \begin{array}{ll}
               \text{HS}(t)  & =  1+(1+\mu_1\mu_5)t^2+\\
                 & +(q\mu_3+q^{-1}\mu_3+1+\mu_2\mu_4+\mu_1\mu_5+\mu_1^2\mu_5^2)t^4+\\
                 & + (1+(q+q^{-1})\mu_3+\dots)t^6+\\
                 & + (1+\mu_2\mu_4+\mu_1\mu_5+\mu_3^2+\dots)t^8
            \end{array}
        \end{equation}
       as obtained from Tab.~\ref{tab:region1HWG}. The Plethystic logarithm, up to syzygies, is:
         \begin{equation}
            \begin{array}{ll}
                \text{PL}(t) & = (1+\mu_1\mu_5)t^2+ \\
                 & +(q\mu_3+q^{-1}\mu_3-1-\mu_1\mu_5)t^4+\\
                & - (q+q^{-1})(\mu_4\mu_5+\mu_3+\mu_1\mu_2)t^6 +\\
                 & -(1+\mu_1\mu_5(1+q^{-2}+q^2)+\mu_2\mu_4-\dots)t^8+\dots \\
            \end{array}
        \end{equation}
         hence the generators are reported in Tab.~\ref{tab:SU(5)1/2+5gens} and the relations in Tab.~\ref{tab:SU(5)1/2+5rel}.
        \renewcommand{\arraystretch}{1.3}
        \begin{table}[h]
            \centering
            \begin{tabular}{|c|c|c|c|} \hline 
                & R-charge & $SU(N_f)$ & $U(1)_I$ \\ \hline \hline
               $M_i^j$  & 2 &  $\mu_1\mu_5$ & 0 \\ \hline
               $S$ & 2 & 1 & 0 \\ \hline
               $I_{ijk}$ & 4 & $\mu_3$ & $1$\\ \hline
               $\tilde{I}_{ijk}$ & 4 & $\mu_3$ & $-1$\\ \hline
            \end{tabular}
            \caption{Generators of the instantonic phase of $SU(4)_{0}+6$ at strong coupling.}
            \label{tab:SU(5)1/2+5gens}
        \end{table}
        \begin{table}[h]
            \centering
            \begin{tabular}{|c|c|c|c|} \hline
                 & R-charge & $SU(N_f)$ & $U(1)_I$  \\ \hline \hline
                $M_i^jM_j^k=S^2\delta_i^k$ & 4 & $\mu_1\mu_5+1$ & 0 \\ \hline
                $I_{jil}M_{r}^{j}=I_{irk}S$ & 6 & $\mu_3$ & $1$  \\ \hline
                $I_{[ijk}M_{r]}^{t}\epsilon_{ta_1\dots a_5}=0$ & 6 & $\mu_4\mu_5$ & $ 1$  \\ \hline
                 $I_{ijk}M_{r}^{t}\epsilon_{ta_1\dots a_{5}}\epsilon^{ijka_1a_2a_3}=0$ & 6 & $\mu_1\mu_2$ & $ 1$  \\ \hline
                $I_{i[jk}I_{mnp]}=0$ & 8 & $\mu_1\mu_5$ & $ 2$ \\ \hline
                 $I\cdot \tilde{I}=S^4$ & 8 & $1$ & 0 \\ \hline
                  $I_{i[jk} \tilde{I}_{mnp]}=S^3M_{i}^{a_1}\epsilon_{a_1jkmnp}$ & 8 & $\mu_1\mu_5$ & 0 \\ \hline
                    $I_{ij[k} \tilde{I}_{mnp]}=S^2M_{[i}^{a_1}M_{j]}^{a_2}\epsilon_{a_1a_2kmnp}$ & 8 & $\mu_2\mu_4$ & 0 \\ \hline
                      $I_{ijk} \tilde{I}_{mnp}=S M_{i}^{a_1}M_{j}^{a_2}M_k^{a_3}\epsilon_{a_1a_2a_3mnp}$ & 8 & $\mu_3^2$ & 0 \\ \hline
            \end{tabular}
            \caption{Chiral ring of the instantonic phase of $SU(4)_{0}+6$ at strong coupling. In the last equation, the symmetrization in the triple of indices is assumed. }
    \label{tab:SU(5)1/2+5rel}
        \end{table}

\paragraph{Region 1, Phase II: $k=0$}\label{par:baryonexample}
Let us consider as an example $SU(4)_0+7$.
    The Hilbert series is:
        \renewcommand{\arraystretch}{1.3}
        \begin{equation}
            \begin{array}{ll}
               \text{HS}(t)  & =  1+(1+\mu_1\mu_6)t^2+\\                 
                 &+(1+\mu_1\mu_6+q^{-1}\mu_3+q\mu_4+\mu_1^2\mu_4^2+\mu_2\mu_3)t^4+ \\
                 & + (q^{-1}\mu_3+q\mu_4+\dots)t^6+\\
                 & + (1+\mu_1\mu_6+\mu_2\mu_5+\mu_3\mu_4\dots)t^8
            \end{array}
        \end{equation}
        as obtained from Tab.~\ref{tab:region1HWG}. The Plethystic logarithm is:
        \renewcommand{\arraystretch}{1.3}
        \begin{equation}
            \begin{array}{ll}
                \text{PL}(t) & = (1+\mu_1\mu_6)t^2+ \\
                 & +(-1-\mu_1\mu_6+q^{-1}\mu_3+q\mu_4)t^4+\\& - (q^{-1}(\mu_3+\mu_4\mu_6+\mu_1\mu_2)+q(\mu_4+\mu_1\mu_3+\mu_5\mu_6))t^6 +\\
                 & -(1+\mu_3\mu_4+q^{-2}\mu_1\mu_5+q^{2}\mu_2\mu_6+\mu_1\mu_5+\mu_1\mu_6)t^8+\dots \\
            \end{array}
        \end{equation}
        and the generators are reported in Tab.~\ref{tab:genphase2SU(3)5} and the relations in Tab.~\ref{tab:relphase2SU(3)5}.
        \renewcommand{\arraystretch}{1.3}
        \begin{table}[h]
            \centering
            \begin{tabular}{|c|c|c|c|} \hline 
             & R-charge & $SU(N_f)$ & $U(1)_B$ \\ \hline \hline
               $M_i^j$  & 2 &  $\mu_1\mu_6$ & 0 \\ \hline
               $S$ & 2 & 1 & 0 \\ \hline
                   $B_{ijkm}$ & 4 & $\mu_4$ & $1$\\ \hline
               $\tilde{B}_{ijk}$ & 4 & $\mu_3$ & $-1$\\ \hline
            \end{tabular}
            \caption{Chiral ring generators for the baryonic phase of $SU(4)_0+7$.}
            \label{tab:genphase2SU(3)5}
        \end{table}
        \begin{table}[h]
            \centering
            \begin{tabular}{|c|c|c|c|} \hline
                 & R-charge & $SU(N_f)$ & $U(1)_B$  \\ \hline \hline
                $M_i^jM_j^k=S^2\delta_i^k$ & 4 & $\mu_1\mu_6+1$ & 0 \\ \hline
                $B_{i_1[i_2\dots i_4}M_{i_5]}^{i_1}=B_{i_2\dots i_5}S$ & 6 & $\mu_4$ & $ 1$  \\ \hline
                $B_{[i_1\dots i_4}M_{i_5]}^{t}\epsilon_{ta_1\dots a_6}=0$ & 6 & $\mu_5\mu_6$ & $ 1$  \\ \hline
                 $B_{i_1\dots i_4}M_{i_5}^{t}\epsilon_{tj_1\dots j_6}\epsilon^{i_1 \dots i_4j_1j_2j_3}=0$ & 6 & $\mu_1\mu_3$ & $ 1$  \\ \hline
                 $\tilde{B}_{i_1i_2i_3} B_{i_4\dots i_7}=SM_{[i_1}^{a_1}M_{i_2}^{a_2}M_{i_3]}^{a_3}\epsilon_{a_1a_2a_3i_4\dots i_7}$ & 8 & $\mu_3\mu_4$ & 0 \\ \hline
                $\tilde{B}_{i_1i_2[i_3} B_{i_4\dots i_7]}=S^2M_{[i_1}^{a_1}M_{i_2]}^{a_2}\epsilon_{a_1a_2i_3\dots i_7}$ & 8 & $\mu_2\mu_5$ & 0 \\ \hline
             $\tilde{B}_{i_1[i_2i_3}B_{i_4\dots i_7]}=S^3M_{i_1}^{a_1}\epsilon_{a_1i_2\dots i_7}$ & 8 & $\mu_1\mu_6$ & 0 \\ \hline
                   $\tilde{B}_{i_1i_2i_3} B_{i_1\dots i_7}\epsilon^{i_1\dots i_7}=S^4$ & 8 & $1$ & 0 \\ \hline
                 $B_{i_1i_2[i_3i_4}B_{i_5\dots i_8]}=0$ & 8 & $\mu_2\mu_6$ & 2 \\ \hline
            \end{tabular}
            \caption{Chiral ring relations for the baryonic phase of $SU(4)_0+7$.}
            \label{tab:relphase2SU(3)5}
        \end{table}

\subsection*{Region 2: $k=0$}\label{par:secondexample}    Let us consider as an example $SU(3)_0+6$, whose Hilbert series:
     \renewcommand{\arraystretch}{1.3}
    \begin{equation}
        \begin{array}{ll}
          \text{HS}(t)   &= 1+(\rho^2+\nu^2+\mu_1\mu_5)t^2+  \\
             & +\rho \nu\mu_3t^3+\\
             & + (1+(\nu^2+\rho^2)\mu_1\mu_5+\dots)t^4+\\
             & + (\nu\rho\mu_3+\dots)t^5\\
             &+ (\nu^2+\rho^2+\nu^2\rho^2\mu_1\mu_5+(\nu^2+\rho^2)\mu_2\mu_4+\mu_3^2+\dots)t^6+\dots
        \end{array}
    \end{equation}
   as obtained from Tab.~\ref{tab:region2HWG}. The Plethystic logarithm is:
    \begin{equation}
        \begin{array}{ll}
           \text{PL}(t)  &= (\mu_1\mu_5+\nu^2+\rho^2)t^2+  \\
             & +\nu\rho\mu_3t^3+\\
             & -(2+\mu_1\mu_5)t^4+\\
             & -(\nu\rho\mu_1\mu_2+2\nu\rho\mu_3+\nu\rho\mu_4\mu_5)t^5+\\
             & -(\nu^2+\rho^2+\mu_3^2+(\nu^2+\rho^2)\mu_2\mu_4+\nu^2\rho^2\mu_1\mu_5)t^6+\dots
        \end{array}
    \end{equation}
    The generators are reported in Tab.~\ref{tab:region2examplegen} and the relations in Tab.~\ref{tab:region2example}.
      \renewcommand{\arraystretch}{1.3}
        \begin{table}[h]
            \centering
            \begin{tabular}{|c|c|c|c|c|} \hline 
                & R-charge & $SU(N_f)$ & $SU(2)_{\nu}$ & $SU(2)_{\rho}$  \\ \hline \hline
               $M_i^j$  & 2 &  $\mu_1\mu_5$ & 1 & 1 \\ \hline
               $S^{(ab)}$ & 2 & 1 & $\nu^2$ & 1 \\ \hline
                $S^{(\dot{a}\dot{b})}$ & 2 & 1 & 1 & $\rho^2$ \\ \hline
               $B_{ijk}^{a,\dot{a}}$ & 3 & $\mu_3$ & $\nu$ & $\rho$\\ \hline
            \end{tabular}
            \caption{Chiral ring generators for strongly coupled $SU(3)_0+6$.}
            \label{tab:region2examplegen}
        \end{table}

          \renewcommand{\arraystretch}{1.3}
        \begin{table}[h]
            \centering
            \begin{tabular}{|c|c|c|c|c|} \hline 
                & R-charge & $SU(N_f)$ & $SU(2)_{\nu}$ & $SU(2)_{\rho}$  \\ \hline \hline
               $M_i^jM_j^k=\delta_i^k(S^{ab})^2=\delta_i^k(S^{\dot{a}\dot{b}})^2$ & 4 & $\mu_1\mu_5+2$ & 1 & 1 \\ \hline
               $M_{[i}^jB_{klm]}^{a,\dot{a}}\epsilon_{ja_1\dots a_5}=0$ & 5 & $\mu_4\mu_5$ & $\nu$ & $\rho$ \\ \hline
               $M_{i}^jB_{klm}^{a,\dot{a}}\epsilon_{jp_1\dots p_5}\epsilon^{klmp_1p_2p_3}=0$ & 5 & $\mu_1\mu_2$ & $\nu$ & $\rho$ \\ \hline
               $M_{[i}^jB_{kl]j}^{a,\dot{a}}=B_{ikl}^{b,\dot{a}}S^{ab}=B_{ikl}^{a,\dot{b}}S^{\dot{a}\dot{b}}=0$ & 5 & $2\mu_3$ & $\nu$ & $\rho$ \\ \hline
               $B_{ijk}^{a,\dot{a}}B_{lmn}^{b,\dot{a}}\epsilon^{ijklmn}=S^{ab}\text{tr}M^2$ & 6 & 1 & $\nu^2$ & 1 \\ \hline
               $B_{ijk}^{a,\dot{a}}B_{lmn}^{a,\dot{b}}\epsilon^{ijklmn}=S^{\dot{a}\dot{b}}\text{tr}M^2$ & 6 & 1 & 1 & $\rho^2$ \\ \hline
               $B_{ijk}^{a,\dot{a}}B_{lmn}^{a,\dot{a}}=M_{[i}^{a_1}M_j^{a_2}M_{k]}^{a_3}\epsilon_{a_1a_2a_3lmn}$ & 6 & $\mu_3^2$ & 1 & 1 \\ \hline
               $B_{ijk}^{a\dot{a}}B_{mnr}^{b\dot{b}}\epsilon^{jkmnrp_1}+S^{ab}S^{\dot{a}\dot{b}}M_i^{p_1}=0$ & 6 & $\mu_1\mu_5$ & $\nu^2$ & $\rho^2$ \\ \hline
               $B_{ij[k}^{a\dot{a}}B_{mln]}^{b\dot{a}}+S^{ab}M_i^{p_1}M_j^{p_2}\epsilon_{p_1p_2kmln}=0$ & 6 & $\mu_2\mu_4$ & $\nu^2$ &  1 \\ \hline 
               $B_{ij[k}^{a\dot{a}}B_{mln]}^{a\dot{b}}+S^{\dot{a}\dot{b}}M_i^{p_1}M_j^{p_2}\epsilon_{p_1p_2kmln}=0$ & 6 & $\mu_2\mu_4$ & 1 &  $\rho^2$ \\ \hline 
            \end{tabular}
            \caption{Chiral ring relations for strongly coupled $SU(3)_0+6$.}
            \label{tab:region2example}
        \end{table}

\subsection*{Region 2: $k=1/2$}\label{sec:1/2example}
  As an example, let us consider $SU(3)_{1/2}+5$. The Hilbert series is:
        \begin{equation}
            \begin{array}{ll}
                \text{HS}(t) & =1+(\mu_1\mu_4+1+\nu^2)t^2+\\
                & +(q^{-1}\nu \mu_3+q\nu \mu_2)t^3+ \\
                & +(1+\mu_1\mu_4)t^4+ \\
                & +(q^{-1}\nu\mu_3+q\nu \mu_2+\dots)t^5 + \\
                 & +(1+\nu^2\mu_2\mu_3+\mu_2\mu_3+\nu^2\mu_1\mu_4+\dots)t^6
            \end{array}
        \end{equation}
        and the plethystic logarithm is:
        \begin{equation}
            \begin{array}{ll}
                \text{PL}(t) & =  (\mu_1\mu_4+1+\nu^2)t^2+\\
                 & +(q^{-1}\nu \mu_3+q\nu \mu_2)t^3+ \\
                 &- (2+\mu_1\mu_4)t^4+\\
                 & -(q\nu\mu_3\mu_4+2q\nu\mu_2+q\nu\mu_1^2+q^{-1}\nu\mu_4^2+2q^{-1}\nu\mu_3+q^{-1}\nu \mu_1\mu_2)t^5\\
                 & -(1+\nu^2+q^2\nu^2\mu_4+q^{-2}\mu_2\mu_4+\mu_2\mu_3+\nu^2\mu_2\mu_3 + q^{-2}\nu^2\mu_1+\nu^2\mu_1\mu_4+q^2\mu_1\mu_3)t^6+\dots
            \end{array}
        \end{equation}
 The generators are reported in Tab.~\ref{tab:SU(3)1/2+5Fgen} and the relations in Tab.~\ref{tab:SU(3)1/2+5Frel}.
          \renewcommand{\arraystretch}{1.3}
        \begin{table}[h]
            \centering
            \begin{tabular}{|c|c|c|c|c|} \hline 
                & R-charge & $SU(N_f)$ & $SU(2)$ & $U(1)$  \\ \hline \hline
               $M_i^j$  & 2 &  $\mu_1\mu_4$ & 1 & 0 \\ \hline
               $S$ & 2 & 1 & 1 & 0 \\ \hline
                $T^{\alpha\beta}$ & 2 & 1 &   $\nu^2$ & 0 \\ \hline
               $B_{ij}^{\alpha}$ & 3 & $\mu_2$ & $\nu$ &1 \\ \hline
                $\tilde{B}_{ijk}^{\alpha}$ & 3 & $\mu_3$ & $\nu$ & $-1$\\ \hline
            \end{tabular}
            \caption{Chiral ring generators of $SU(3)_{1/2}+5F$ at strong coupling.}
            \label{tab:SU(3)1/2+5Fgen}
        \end{table}
       
          \renewcommand{\arraystretch}{1.4}
         \begin{table}[h]
            \centering
            \begin{tabular}{|c|c|c|c|c|} \hline 
                & R-charge & $SU(N_f)$ & $SU(2)$ & $U(1)$  \\ \hline \hline
            $M_i^jM_j^k=(S^2)\delta_i^k$ & 4 & $\mu_1\mu_4+1$ & 1 & 0 \\ \hline
              $S^2=T^2$ & 4 & $1$ & 1 & 0 \\ \hline
            $B^{\alpha}_{[ij}M^l_{k]}\epsilon_{la_1\dots a_4}=0$ & 5 & $\mu_3\mu_4$ & $\nu$ & 1 \\ \hline
  $SB^{\alpha}_{ij}+B^{\alpha}_{ik}M^k_j=0$ & 5 & $\mu_2+\mu_1^2$ & $\nu$ & 1 \\ \hline
              $B^{\beta}_{ij}T^{\beta\alpha}+B^{\alpha}_{k[i}M^k_{j]}=0$ & 5 & $\mu_2$ & $\nu$ & 1 \\ \hline\makecell{$B_{ij}^{\alpha}\tilde{B}^{\beta}_{klm}=(\epsilon^{\alpha\beta}S^3+S^2T^{\alpha\beta})\epsilon_{ijklm}+$\\$+(T^{\alpha\beta}+\epsilon^{\alpha\beta}S)M_i^pM_j^t\epsilon_{ptklm}$\\$+ST^{\alpha\beta}M_i^p\epsilon_{pjklm}$} & 6 & $1+\mu_2\mu_3+\mu_1\mu_4$ &  $\nu^2+1$ & 0 \\ \hline
                  $B^{\alpha}_{[ij}B^{\beta}_{kl]}=0$ & 6 & $\mu_4$ & $\nu^2$ & 2 \\ \hline
                   $B^{\alpha}_{i[j}B^{\alpha}_{kl]}=0$ & 6 & $\mu_1\mu_3$ & $1$ & 2 \\ \hline
            \end{tabular} 
            \caption{Chiral ring relations for $SU(3)_{1/2}+5F$ at strong coupling.}
            \label{tab:SU(3)1/2+5Frel}
        \end{table} 

\subsection*{Region 3: $k=0$}\label{sec:0example}
Let us consider $SU(3)_0+8F$. The Hilbert series is:
 \renewcommand{\arraystretch}{1.2}
    \begin{equation}
        \begin{array}{ll}
            \text{HS}(t) & =1+(\mu_1\mu_9)t^2+ \\
            & + \mu_4t^3+\\
            &  +(\mu_5+\mu_1\mu_5\mu_9)t^5+\\
             & +(\mu_3\mu_7+\dots)t^6+\dots \\
        \end{array}
    \end{equation}
    and the plethystic logarithm is:
     \renewcommand{\arraystretch}{1.2}
    \begin{equation}
        \begin{array}{ll}
            \text{PL}(t) & =(\mu_1\mu_9)t^2+\\
            & + \mu_4 t^3+ \\
            & -\mu_1\mu_7t^4 + \\
            & - (\mu_6\mu_9+\mu_1\mu_4)t^5+ \\
            &-\mu_3\mu_7 t^6+\dots\\
        \end{array}
    \end{equation}
   The generators are reported in Tab.~\ref{tab:Nc=3,k=0,Nf=8gen} and the relations in Tab.~\ref{tab:tab:Nc=2,k=0,Nf=8rel}.
     \renewcommand{\arraystretch}{1.3}
        \begin{table}[h]
            \centering
            \begin{tabular}{|c|c|c|} \hline 
                & R-charge  & $SU(10)$ \\ \hline \hline
               $M_i^j$  & 2 &  $\mu_1\mu_9$ \\ \hline
               $I_{a_1\dots a_5}$ & 3 & $\mu_5$ \\ \hline
            \end{tabular}
            \caption{Chiral ring generators of strongly coupled $SU(3)_0+8F$.}
            \label{tab:Nc=3,k=0,Nf=8gen}
        \end{table}
          \renewcommand{\arraystretch}{1.3}
        \begin{table}[h]
            \centering
            \begin{tabular}{|c|c|c|} \hline 
                & R-charge  & $SU(10)$ \\ \hline \hline
             $M_i^jM_j^k-\text{tr}M^2\delta_i^k/10=0$ & 4 & $\mu_1\mu_9$ \\ \hline
             $M_{i_1}^{j}I_{i_2\dots i_5]}\epsilon_{jp_1\dots p_{9}}\epsilon^{i_2\dots i_5 p_1 \dots p_5}=0$ & 5 & $\mu_1 \mu_4$ \\ \hline
             $M_{[i_1}^jI_{i_2\dots i_5]}\epsilon_{j a_1\dots a_9}=0$ & 5 & $\mu_6\mu_9$ \\ \hline
             $I_{i_1\dots i_3[i_4i_5}I_{i_6\dots i_{10}]}=M_{[i_1}^{a_1}M_{i_2}^{a_2}M_{i_3]}^{a_3}\epsilon_{a_1a_2i_3 \dots i_7}$ & 6 & $\mu_3 \mu_7$ \\ \hline
            \end{tabular}
            \caption{Chiral ring relations of strongly coupled $SU(3)_0+8F$.}
            \label{tab:tab:Nc=2,k=0,Nf=8rel}
        \end{table}

\subsection*{Region 3: $k=1/2$}\label{sec:region31/2}
Let us consider $SU(2)_{1/2}+5F$. The Hilbert series is:
  \renewcommand{\arraystretch}{1.2}
        \begin{equation}
            \begin{array}{ll}
                \text{HS}(t) & =1+(\nu^2+\mu_1\mu_5)t^2 \\
                 & +\nu\mu_3 t^3+\\
                 &+(1+\dots )t^4+\\
                 & +(\nu\mu_3+\dots)t^5+\\
                 & +(\dots)t^6
            \end{array}
        \end{equation}
        and the plethystic logarithm is:
        \begin{equation}
            \begin{array}{ll}
                \text{PL}(t) & =(\nu^2+\mu_1\mu_5)t^2+ \\
                 & +\nu\mu_3 t^3+\\
                 & -(1+\mu_1\mu_5)t^4+\\
                 & -\nu(\mu_4\mu_5+\mu_3+\mu_1\mu_2)t^5+\\
                 & -(1+\mu_2\mu_4+\nu^2\mu_1\mu_5)t^6
            \end{array}
        \end{equation}
The generators are reported in Tab.~\ref{tab:N_c=2,k=1/2,N_f=5gen} and the relations in Tab.~\ref{tab:N_c=2,k=1/2,N_f=5rel}.
        \renewcommand{\arraystretch}{1.3}
        \begin{table}[h]
            \centering
            \begin{tabular}{|c|c|c|c|} \hline 
                & R-charge  & $SU(6)$ & $SU(2)$ \\ \hline \hline
               $M_i^j$  & 2 &  $\mu_1\mu_5$ & 1 \\ \hline
               $S^{(ab)}$ & 2 & 1 & $\nu^2$ \\ \hline
               $B_{i_1i_2i_3}$ & 3 & $\mu_3$ & $\nu$ \\ \hline
            \end{tabular}
            \caption{Chiral ring generators of $SU(2)_{1/2}+5F$ at strong coupling.}
            \label{tab:N_c=2,k=1/2,N_f=5gen}
        \end{table}
          \renewcommand{\arraystretch}{1.3}
        \begin{table}[h]
            \centering
            \begin{tabular}{|c|c|c|c|} \hline 
              Constraints  & R-charge  & $SU(6)$ & $SU(2)$ \\ \hline \hline
               $M_i^jM_j^k=S^2\delta_i^k$  & 4 &  $\mu_1\mu_5+1$ & 1 \\ \hline
              $B_{j[i_1i_2}^aM^j_{i_3]}=B^b_{i_1i_2i_3}S^{ba}$ & 5 & $\mu_3$ & $\nu$ \\ \hline
              $B_{[i_1i_2i_3}^aM_{i_4]}^l\epsilon_{lp_1\dots p_5}=0$ & 5 & $\mu_4\mu_5$ & $\nu$ \\ \hline
              $B_{i_1i_2i_3}^aM_k^l\epsilon_{lp_1\dots p_4}\epsilon^{i_1i_2i_3p_1p_2}=0$ & 5 & $\mu_1\mu_2$ & $\nu$ \\ \hline
             $B^a_{[i_1i_2i_3}B^a_{i_4i_5i_6]}=0$ & 6 & 1 & 1 \\ \hline
              $B^{(a}_{i_1[i_2i_3}B^{b)}_{i_4i_5i_6]}=0$ & 6 & $\mu_1\mu_5$ & $\nu^2$ \\ \hline
               $B^{a}_{i_1i_2[i_3}B^{a}_{i_4i_5i_6]}=0$ & 6 &  $\mu_2\mu_4$ & 1 \\ \hline
            \end{tabular}
            \caption{Chiral ring constraints of $SU(2)_{1/2}+5F$ at strong coupling.}
            \label{tab:N_c=2,k=1/2,N_f=5rel}
        \end{table}

\subsection*{Region 3: $k \geq 1$}\label{sec:region3greater}
let us consider $SU(2)_{2}+2F$. The Hilbert series is:
        \begin{equation}
            \begin{array}{ll}
                \text{HS}(t) & =1+(1+\mu_1\mu_2)t^2+  \\
                 & +(1+q\mu_2+q^{-1}\mu_1+\mu_1\mu_2+\mu_1^2\mu_2^2)t^4+\\
                 & + (q\mu_2+q^{-1}\mu_1+\dots)t^6
            \end{array}
        \end{equation}
        and the plethystic logarithm is:
        \begin{equation}
            \begin{array}{ll}
               \text{PL}(t)  &  = (1+\mu_1\mu_2)t^2+\\
                 & +(q\mu_2+q^{-1}\mu_1-1-\mu_1\mu_2)t^4+\\
                 & -(q\mu_2+q^{-1}\mu_1+q\mu_1^2+q^{-1}\mu_2^2)t^6 + \\
                 & -(1+\mu_1\mu_2)t^8+\dots
            \end{array}
        \end{equation}
       The generators are reported in Tab.~\ref{tab:N_c=2,k=2,N_f=2gen} and the relations in Tab.~\ref{tab:N_c=2,k=2,N_f=2rel}.
         \renewcommand{\arraystretch}{1.3}
        \begin{table}[h]
            \centering
            \begin{tabular}{|c|c|c|c|} \hline 
                & R-charge  & $SU(3)$ & $U(1)$ \\ \hline \hline
               $M_i^j$  & 2 &  $\mu_1\mu_2$ & 0 \\ \hline
               $S$ & 2 & 1 & 0 \\ \hline
                $I_{ij}$ & 4 & $\mu_1$ & $1$ \\ \hline
               $\tilde{I}_{i}$ & 4 & $\mu_2$ & $-1$ \\ \hline
            \end{tabular}
            \caption{Chiral ring generators of $SU(2)_{2}+2F$ at strong coupling.}
            \label{tab:N_c=2,k=2,N_f=2gen}
        \end{table}
         \begin{table}[h]
            \centering
            \begin{tabular}{|c|c|c|c|} \hline 
                & R-charge  & $SU(3)$ & $U(1)$ \\ \hline \hline
               $M_i^jM_j^k=S^2\delta_i^k$  & 4 &  $\mu_1\mu_2+1$ & 0 \\ \hline
               $I_{i[j}M^i_{k]}=SI_{jk}$ & 6 & $\mu_2$ & $1$ \\ \hline
               $I_{i(j}M^{i}_{k)}=0$ & 6 & $\mu_1^2$ & 1 \\ \hline
               $\tilde{I}_{[i_1}I_{i_2i_3]}=S^4 \epsilon_{i_1i_2i_3}$ & 8 & 1 & 0 \\ \hline
               $\tilde{I}_{[i_1}I_{i_2]i_3}=S^3M_{i_3}^j\epsilon_{ji_1i_2}$ & 8 & $\mu_1\mu_2$ & 0 \\ \hline
            \end{tabular}
            \caption{Chiral ring constraints of $SU(2)_{2}+2F$ at strong coupling.}
            \label{tab:N_c=2,k=2,N_f=2rel}
        \end{table}

\subsection*{SO: no symmetry enhancement}\label{par:SO(8)}
Let us consider SO(8); the pure theory is $A_5$, the theory with 1 flavour is the same as SO(6) with 1 flavour except the R-charge is now 6. The theory with 2 flavours is the same as $SO(7)+2F$, with changed R-charge. For 3 flavours, the generators are reported in Tab.~\ref{tab:SO(8)+3gen} and the relations in Tab.~\ref{tab:SO(8)+3Frel}.
\begin{table}[h]
        \centering
        \begin{tabular}{|c|c|c|c|} \hline
          Generators   & R-charge &  $Sp(3)$ & $U_I(1)$ \\ \hline \hline
            $M_{(ij)}$ & 2 & $\mu_1^2$ & 0 \\ \hline
            $S$ & 2 & 1 & 0 \\ \hline
            $I_{[ijk]},\tilde{I}_{[ijk]}$ & 6 & $\mu_3$ & $\pm 1$ \\ \hline
        \end{tabular}
        \caption{Chiral ring generators of $SO(8)+3F$ at strong coupling.}
        \label{tab:SO(8)+3gen}
    \end{table}
\begin{table}[h]
    \centering
    \begin{tabular}{|c|c|c|c|} \hline
       Constraints  & R-charge & $Sp(3)$ & $U_I(1)$ \\ \hline \hline
       $M^2=S^2$  & 4 & $\mu_2+1$ & 0 \\ \hline
       $MI=SI$ & 8 & $\mu_3+\mu_1\mu_2$ & $\pm 1$ \\ \hline
       $(I\tilde{I})_{1}=S^6$ & 12 & 1 & 0 \\ \hline
        $(I\tilde{I})_{\mu_1^2}=S^5M$ & 12 & $\mu_1^2$ & 0 \\ \hline
    $(I\tilde{I})_{\mu_2^2}=S^4\wedge^3M$ & 12 & $\mu_2^2$ & 0 \\\hline
     $(I\tilde{I})_{\mu_3^2}=S\wedge^5M$ & 12 & $\mu_3^2$ & 0 \\\hline
       $(I \cdot I)_{\mu_1^2}=0$ & 12 & $\mu_1^2$ & $\pm 2$ \\ \hline
    \end{tabular}
    \caption{Chiral ring relations of $SO(8)+3F$ at strong coupling. While we notice the appearance of the expected relations at order $12$, the syzigies make it quite complex the identification.}
    \label{tab:SO(8)+3Frel}
\end{table}

\subsection*{SO: partial symmetry enhancement}\label{par:exSO}
Let us consider $SO(6)$ with $2$ flavours, whose Hilbert series is:
\begin{equation}
    \begin{array}{ll}
       \text{HS}(t)  & =1+(\mu_1^2+\nu^2)t^2+ \\
         & +(\nu^2+1+\dots)t^4+\\
         & + (\nu^2\mu_2+\dots)t^6+\\
         & + (1+\nu^4+\dots)t^8
    \end{array}
\end{equation}
and the plethystic logarithm is:
\begin{equation}
    \begin{array}{ll}
        \text{PL}(t) & =  (\mu_1^2+\nu^2)t^2+\\
         & +(\nu^2\mu_2-\mu_2-1)t^4+ \\
         & -(\nu^2\mu_2-\nu^2\mu_1^2)t^6+ \\
         & -(\nu^4+1)t^8+\dots
    \end{array}
\end{equation}
and generators are:
\begin{table}[H]
        \centering
        \begin{tabular}{|c|c|c|c|} \hline
          Generators   & R-charge &  $Sp(2)$ & $SU(2)$ \\ \hline
            $M_{(ij)}$ & 2 & $\mu_1^2$ & 1 \\
            $S^{ab}$ & 2 & 1 & $\nu^2$ \\
            $I_{[i_1i_2]}^{(ab)}$ & 4 & $\mu_2$ & $\nu^2$ \\ \hline
        \end{tabular}
        \caption{Chiral ring generators of strongly coupled $SO(6)+2F$.}
        \label{tab:SO(6)+2gen}
\end{table}
and the constraints are:
\begin{table}[H]
        \centering
        \begin{tabular}{|c|c|c|c|} \hline
          Generators   & R-charge &  $Sp(2)$ & $SU(2)$ \\ \hline \hline
            $M_{ij}M_{kl}\Omega^{jk}=S^2\Omega_{il}$ & 4 & $\mu_2+1$ & 1 \\ \hline
            $M_{kl}I^{ab}_{ij}\Omega^{ki}+S^{ab}I^{ba}_{ij}=0$ & $6$ & $\mu_1^2+\mu_2$ & $\nu^2$ \\ \hline
            $I_{ij}^{ab}I_{ij}^{ab}=S^4$ & $2N_c-4$ & $1$ & $1$ \\ \hline
              $I_{i_1\dots i_{N_f}}^{(a_1a_2}I_{i_1 \dots i_{N_f}}^{a_3a_4)}=S^{(a_1a_2}S^{a_3a_4)}\text{tr}S^{2}$ & $2N_c-4$ & $1$ & $\nu^{4}$ \\ \hline
        \end{tabular}
        \caption{Chiral ring relations of $SO(6)+2F$.}
        \label{tab:SO(6)+2rel}
\end{table}

\newpage

\bibliographystyle{JHEP}
\bibliography{bibli.bib}

@article{tachikawa2015instanton,
  title={Instanton operators and symmetry enhancement in 5d supersymmetric gauge theories},
  author={Tachikawa, Yuji},
  journal={Progress of Theoretical and Experimental Physics},
  volume={2015},
  number={4},
  pages={043B06},
  year={2015},
  publisher={Oxford University Press},
  eprint={1501.01031},
  archivePrefix={arXiv},
  primaryClass={hep-th}
}

@article{Gaiotto:2015una,
    author = "Gaiotto, Davide and Kim, Hee-Cheol",
    title = "{Duality walls and defects in 5d $ \mathcal{N}=1 $ theories}",
    eprint = "1506.03871",
    archivePrefix = "arXiv",
    primaryClass = "hep-th",
    doi = "10.1007/JHEP01(2017)019",
    journal = "JHEP",
    volume = "01",
    pages = "019",
    year = "2017"
}

@article{Argyres:1996eh,
    author = "Argyres, Philip C. and Plesser, M. Ronen and Seiberg, Nathan",
    title = "{The Moduli space of vacua of N=2 SUSY QCD and duality in N=1 SUSY QCD}",
    eprint = "hep-th/9603042",
    archivePrefix = "arXiv",
    reportNumber = "RU-96-07, WIS-96-1-PH",
    doi = "10.1016/0550-3213(96)00210-6",
    journal = "Nucl. Phys. B",
    volume = "471",
    pages = "159--194",
    year = "1996"
}

@inproceedings{joseph1976minimal,
  title={The minimal orbit in a simple Lie algebra and its associated maximal ideal},
  author={Joseph, Anthony},
  doi ="10.24033/asens.1302",
  booktitle={Annales scientifiques de l'{\'E}cole normale sup{\'e}rieure},
  volume={9},
  number={1},
  pages={1--29},
  year={1976}
}

@article{Feng:2007ur,
    author = "Feng, Bo and Hanany, Amihay and He, Yang-Hui",
    title = "{Counting gauge invariants: The Plethystic program}",
    eprint = "hep-th/0701063",
    archivePrefix = "arXiv",
    doi = "10.1088/1126-6708/2007/03/090",
    journal = "JHEP",
    volume = "03",
    pages = "090",
    year = "2007"
}

@misc{EVnew,
    author       = {Van den Driessche, Elias},
    howpublished = {In preparation}
}

@article{Yonekura:2015ksa,
    author = "Yonekura, Kazuya",
    title = "{Instanton operators and symmetry enhancement in 5d supersymmetric quiver gauge theories}",
    eprint = "1505.04743",
    archivePrefix = "arXiv",
    primaryClass = "hep-th",
    doi = "10.1007/JHEP07(2015)167",
    journal = "JHEP",
    volume = "07",
    pages = "167",
    year = "2015"
}

@article{Jefferson:2017ahm,
    author = "Jefferson, Patrick and Kim, Hee-Cheol and Vafa, Cumrun and Zafrir, Gabi",
    title = "{Towards classification of 5d SCFTs: Single gauge node}",
    eprint = "1705.05836",
    archivePrefix = "arXiv",
    primaryClass = "hep-th",
    doi = "10.21468/SciPostPhys.14.5.122",
    journal = "SciPost Phys.",
    volume = "14",
    number = "5",
    pages = "122",
    year = "2023"
}

@article{Gorsky:2000px,
    author = "Gorsky, A. and Mironov, A.",
    editor = "Aratyn, H. and Sorin, Alexander Savelievich",
    title = "{Integrable many body systems and gauge theories}",
    eprint = "hep-th/0011197",
    archivePrefix = "arXiv",
    reportNumber = "FIAN-TD-30-00, ITEP-TH-64-00",
    pages = "33--176",
    month = "11",
    year = "2000"
}

@article{Chaban:2025mjz,
    author = "Chaban, Igor and Nekrasov, Nikita",
    title = "{Lectures on Gauge theories and Many-Body systems}",
    eprint = "2512.23099",
    archivePrefix = "arXiv",
    primaryClass = "math-ph",
    month = "12",
    year = "2025"
}

@article{Gray:2008yu,
    author = "Gray, James and Hanany, Amihay and He, Yang-Hui and Jejjala, Vishnu and Mekareeya, Noppadol",
    title = "{SQCD: A Geometric Apercu}",
    eprint = "0803.4257",
    archivePrefix = "arXiv",
    primaryClass = "hep-th",
    reportNumber = "IHES-P-08-04, IMPERIAL-TP-08-AH-03",
    doi = "10.1088/1126-6708/2008/05/099",
    journal = "JHEP",
    volume = "05",
    pages = "099",
    year = "2008"
}

@phdthesis{Zhong:2022pmg,
    author = "Zhong, Zhenghao",
    title = "{Magnetic quivers {\textendash} a new perspective on supersymmetric gauge theories}",
    doi = "10.25560/103461",
    school = "Imperial Coll., London",
    year = "2022"
}

@article{Bourget:2019rtl,
    author = "Bourget, Antoine and Cabrera, Santiago and Grimminger, Julius F. and Hanany, Amihay and Zhong, Zhenghao",
    title = "{Brane Webs and Magnetic Quivers for SQCD}",
    eprint = "1909.00667",
    archivePrefix = "arXiv",
    primaryClass = "hep-th",
    doi = "10.1007/JHEP03(2020)176",
    journal = "JHEP",
    volume = "03",
    pages = "176",
    year = "2020"
}

@article{Akhond:2024nyr,
    author = "Akhond, Mohammad and Arias-Tamargo, Guillermo and Carta, Federico and Grimminger, Julius F. and Hanany, Amihay",
    title = "{On brane systems with O$^{+}$ planes {\textemdash} 5d and 6d SCFTs}",
    eprint = "2411.02491",
    archivePrefix = "arXiv",
    primaryClass = "hep-th",
    doi = "10.1007/JHEP03(2025)137",
    journal = "JHEP",
    volume = "03",
    pages = "137",
    year = "2025"
}

@article{Hanany:2026eyi,
    author = "Hanany, Amihay and Tomasiello, Alessandro and Van den Driessche, Elias",
    title = "{5d Higgs branch and instanton magnetization}",
    eprint = "2603.29679",
    archivePrefix = "arXiv",
    primaryClass = "hep-th",
    reportNumber = "Imperial/TP/26/AH/03",
    month = "3",
    year = "2026"
}

@article{Hanany:2025ctg,
    author = "Hanany, Amihay and Driessche, Elias Van den",
    title = "{Higgs branch of 5d {\ensuremath{\mathscr{N}}} = 1 symplectic gauge theories and dressed instanton operators}",
    eprint = "2507.08669",
    archivePrefix = "arXiv",
    primaryClass = "hep-th",
    reportNumber = "Imperial/TP/25/AH/07",
    doi = "10.1007/JHEP09(2026)108",
    journal = "JHEP",
    volume = "09",
    pages = "108",
    year = "2026"
}

@article{Hanany:2025jwo,
    author       = {Hanany, Amihay and Van den Driessche, Elias},
    title        = {Chiral ring along the {RG} flow in 5d {$\mathscr{N}$} = 1},
    eprint       = {2510.15635},
    archivePrefix = {arXiv},
    primaryClass = {hep-th},
    reportNumber = {Imperial/TP/25/AH/10},
    doi          = {10.1007/JHEP09(2026)096},
    journal      = {JHEP},
    volume       = {09},
    pages        = {096},
    year         = {2026}
}

@article{gledhill2023poisson,
  title={Poisson brackets for some {Coulomb} branches},
  author={Gledhill, Kirsty and Hanany, Amihay},
  journal={Journal of High Energy Physics},
  volume={2023},
  number={3},
  pages={1--56},
  year={2023},
  publisher={Springer},
  eprint={2210.02966},
  archivePrefix={arXiv},
  primaryClass={hep-th}
}

@article{bourget2020magnetic,
  title={Magnetic quivers from brane webs with {O5} planes},
  author={Bourget, Antoine and Grimminger, Julius F and Hanany, Amihay and Sperling, Marcus and Zhong, Zhenghao},
  journal={Journal of High Energy Physics},
  volume={2020},
  number={7},
  pages={1--82},
  year={2020},
  publisher={Springer},
  eprint={2004.04082},
  archivePrefix={arXiv},
  primaryClass={hep-th}
}

@article{hanany2014highest,
  title={Highest weight generating functions for {Hilbert} series},
  author={Hanany, Amihay and Kalveks, Rudolph},
  journal={Journal of High Energy Physics},
  volume={2014},
  number={10},
  pages={1--68},
  year={2014},
  publisher={Springer},
  eprint={1408.4690},
  archivePrefix={arXiv},
  primaryClass={hep-th}
}

@article{beem2017deformation,
  title={Deformation quantization and superconformal symmetry in three dimensions},
  author={Beem, Christopher and Peelaers, Wolfger and Rastelli, Leonardo},
  journal={Communications in Mathematical Physics},
  volume={354},
  number={1},
  pages={345--392},
  year={2017},
  publisher={Springer},
  eprint={1601.05378},
  archivePrefix={arXiv},
  primaryClass={hep-th}
}

@article{cabrera2019tropical,
  title={Tropical geometry and five dimensional {Higgs} branches at infinite coupling},
  author={Cabrera, Santiago and Hanany, Amihay and Yagi, Futoshi},
  journal={Journal of High Energy Physics},
  volume={2019},
  number={1},
  pages={1--50},
  year={2019},
  publisher={Springer},
  eprint={1810.01379},
  archivePrefix={arXiv},
  primaryClass={hep-th}
}

@article{seiberg1996five,
  title={Five dimensional {SUSY} field theories, non-trivial fixed points and string dynamics},
  author={Seiberg, Nathan},
  journal={Physics Letters B},
  volume={388},
  number={4},
  pages={753--760},
  year={1996},
  publisher={Elsevier},
  eprint={hep-th/9608111},
  archivePrefix={arXiv},
  primaryClass={hep-th}
}

@article{zafrir2015instanton,
  title={Instanton operators and symmetry enhancement in 5d supersymmetric {USp}, {SO} and exceptional gauge theories},
  author={Zafrir, Gabi},
  journal={Journal of High Energy Physics},
  volume={2015},
  number={7},
  pages={1--34},
  year={2015},
  publisher={Springer},
  eprint={1503.08136},
  archivePrefix={arXiv},
  primaryClass={hep-th}
}

@article{cremonesi2017instanton,
  title={Instanton operators and the {Higgs} branch at infinite coupling},
  author={Cremonesi, Stefano and Ferlito, Giulia and Hanany, Amihay and Mekareeya, Noppadol},
  journal={Journal of High Energy Physics},
  volume={2017},
  number={4},
  pages={1--43},
  year={2017},
  publisher={Springer},
  eprint={1505.06302},
  archivePrefix={arXiv},
  primaryClass={hep-th}
}

\end{document}